%% file: paper.tex
\RequirePackage{fix-cm}
\documentclass[smallextended,natbib]{svjour3}       % onecolumn (second format)
\smartqed  % flush right qed marks, e.g. at end of proof
\usepackage{graphicx}
\usepackage{lscape}
\usepackage{subcaption}
\usepackage{multirow}
\usepackage{float}
\usepackage{multicol}
\usepackage{siunitx}
\usepackage{adjustbox}
\usepackage{caption}
\usepackage{ifthen}
\usepackage{amssymb}
\usepackage{soul}
\usepackage[table]{xcolor}
\usepackage[hyphens]{url}
\usepackage{hyperref}
\usepackage{graphicx}
\usepackage{tcolorbox}
\usepackage{mdframed}
\usepackage{tikz}

\newboolean{showcomments}
\setboolean{showcomments}{true}
\ifthenelse{\boolean{showcomments}}
 { \newcommand{\mynote}[2]{
      \fbox{\bfseries\sffamily\scriptsize#1}
        {\small$\blacktriangleright$\textsf{\emph{#2}}$\blacktriangleleft$}}}
        { \newcommand{\mynote}[2]{}}

\newcommand{\az}[0]{AndroZoo\xspace}
\newcommand{\vt}[0]{VirusTotal}
\newcommand{\ag}[0]{Androguard}
\newcommand{\iccta}[0]{IccTA}

\newcommand{\ggl}[0]{Google}
\newcommand{\gp}[0]{Google Play}

\newcommand{\rotCell}[4]{\multicolumn{#1}{#2}{\rlap{\rotatebox{#3}{#4}~}}}
\newcommand{\highlight}[1]{\begin{tcolorbox}[leftrule=1mm,rightrule=1mm,toprule=0mm,bottomrule=0mm,left=2pt,right=2pt,top=2pt,bottom=2pt]
#1
\end{tcolorbox}
}

\usepackage{xspace}
\newcommand{\appspkg}{\num{131}}
\newcommand{\appsdate}{\num{184432}}
\newcommand{\totalApps}{\num{184563}}
\newcommand{\appsAfterFilter}{\num{4103}}
\newcommand{\numCovidApps}{\num{35613}}
\newcommand{\appsAfterUniqVersion}{\num{750}}
\newcommand{\covidAppsInAz}{\num{44}}
\newcommand{\additionalApps}{\num{48}}
\newcommand{\covidrelatedApps}{\num{92}} % covidAppsInAz + additionalApps
\newcommand{\covidAppsNotAvailableDesc}{\num{14}}
\newcommand{\covidrelatedAppsWithDesc}{\num{78}} % covidrelatedApps - covidAppsNotAvailableDesc
\newcommand{\countriesNotAvailable}{\num{16}}
\newcommand{\covid}{Covid-19\xspace}
\newcommand{\covidRelated}{Covid-related\xspace}

\begin{document}

\title{A First Look at Android Applications in \gp{} related to Covid-19}
%\subtitle{Subtitle}


\author{Jordan Samhi \and
        Kevin Allix \and
        Tegawendé F. Bissyandé \and
        Jacques Klein
}

\authorrunning{Jordan Samhi et al.} % if too long for running head

\institute{
University of Luxembourg, SnT\\
6 Rue Richard Coudenhove-Kalergi,\\
1359 Luxembourg\\
Tel.: +352 46 66 44 5000\\
\email{firstname.lastname@uni.lu}
}

\date{Received: date / Accepted: date}

\maketitle

\begin{abstract}
\input{abstract}
\end{abstract}

\input{introduction}
\input{background}

\input{experiment_setup}
\input{results}
\input{discussion}
\input{related_work}

\input{conclusion}

\section{Acknowledgment}
This work was supported (1) by the Luxembourg National Fund under the project CHARACTERIZE C17/IS/11693861, the AFR grant 14596679, and (2) by the SPARTA project, which has received funding from the European Union's Horizon 2020 research and innovation program under grant agreement No 830892.

%\begin{acknowledgements}
%\end{acknowledgements}

\bibliographystyle{spbasic}      % basic style, author-year citations
\bibliography{bib}

\appendix

\input{complexity}

\end{document}

%% file: abstract.tex
Due to the convenience of access-on-demand to information and business 
solutions, mobile apps have become an important asset in the digital world. 
In the context of the \covid pandemic, app developers have joined the response 
effort in various ways by releasing apps that target different user bases 
(e.g., all citizens or journalists), offer different services (e.g., location 
tracking or diagnostic-aid), provide generic or specialized information, etc. 
While many apps have raised some concerns by spreading misinformation or even 
malware, the literature does not yet provide a clear landscape  of the 
different apps that were developed. 
In this study, we focus on the Android ecosystem and investigate \covidRelated 
Android apps. In a best-effort scenario, we attempt to systematically identify 
all relevant apps and study their characteristics with the objective to 
provide a first taxonomy of \covidRelated apps, broadening the relevance 
beyond the implementation of contact tracing. Overall, our study yields a 
number of empirical insights that contribute to enlarge the knowledge on 
\covidRelated apps: 
(1) Developer communities contributed rapidly to the 
\covid, with dedicated apps released as early as January 2020;
(2) \covidRelated apps 
deliver digital tools to users (e.g., health diaries), serve to broadcast information to users (e.g., spread statistics), and collect data from users (e.g., for tracing); 
(3) \covidRelated apps are less complex than standard apps; 
(4) they generally do not seem to leak sensitive data; 
(5) in the majority of cases, 
\covidRelated apps are released by entities with past experience on the 
market, mostly official government entities or public health organizations.

%% file: introduction.tex
\section{Introduction} \label{introduction} 

The outbreak of the \covid pandemic in late December 2019 has quickly and severely impacted countries worldwide, becoming one of the most important health crisis of the 21$^{st}$ century~\citep{spinelli2020covid,remuzzi2020covid}. 
The infectious agent, a {\em coronavirus}, identified as responsible for this disease is notoriously 
difficult to pin down: while it leaves many infected people without symptoms, it can lead to a common cold for some and even severe respiratory disorders to others~\citep{clerkin2020covid}. 
So far, the \covid has brought about a human tragedy, with hundreds of thousands of 
lives lost, as well as an economic downturn due to the lockdown of over three billion people (half of humanity)~\citep{mahase2020coronavirus, dudel2020monitoring}. 

The scale of \covid effects has urged stakeholders at different levels (government, local authorities, private companies, academia, NGOs, and citizens) to plan and implement measures for addressing the virus spread. 
In particular, while pharmaceutical research is embarked on a long journey to develop a vaccine, non-pharmaceutical innovations are sought to contribute to responding to the outbreak. 
Chief among the technologies that are employed, Information and Communication Technology has been widely leveraged in various capacities across all regions and targeting broadly all levels of society. 
For example, news and sensitization messages have been viral thanks to the use of internet-based services such as social networks. 
Given the widespread use of handheld devices such as smartphones, users are keen to install 
applications (often referred to as \emph{apps} in the mobile realm) that have specific purposes for entertainment, business, productivity, news, and social networking. 
Following the outbreak of the pandemic, authorities, non-governmental organizations, and independent developers have engaged in an app development race to provide readily-available digital tools to the modern citizen. 

We focus in this paper on the case of the Android ecosystem. With the largest market share on mobiles (86\% in 2020~\citep{idc}), Android constitutes a prime choice for developers and users alike. 
Initial reporting on \covid related apps are focused on the problems that such apps raise: 
(1) \citet{he2020virus} have already explored the case of coronavirus-themed Android malware; 
(2) Google, the maintainer of the Google Play (i.e., the official Android app market) 
has decided to crack down on \covid apps to combat misinformation, sometimes 
with an excessive zeal (i.e., legitimate apps can be temporarily banned just 
for sharing \covid information~\citep{googlecovid,googleupdate}).

Our study is of a broader and more generic dimension. 
It is about {\bf characterizing the applications that are related to the \covid outbreak}: 
What are they for? Who developed them? To what extent can they be considered 
dangerous? These are some of the questions that we undertake to investigate. 
To that end, we have considered \totalApps{} apps released in the time 
window of July 2019-May 2020 and collected in
the \az{} dataset~\citep{Allix:2016:ACM:2901739.2903508}. 
From this initial set of Android apps, by following well-defined heuristics, we were able to identify \covidAppsInAz{} \covidRelated apps.
Given the limitations of the crawling of \az{}, we also considered other online 
resources such as the GitHub search engine, specialized \covid technology 
blogs, etc., and we finally raise the number of collected \covidRelated apps 
to \covidrelatedApps{}.
We extract different features from these 
apps and provide summary statistics on their characterization.

This paper presents our analyses exploring the use of permissions and libraries, the presence of leaks, the  malicious status, the code size and complexity, the authorship, and the described purposes. 
Mainly, we establish that most of the apps are made
for informing people, monitoring their health, and tracing users with the goal of preventing 
the spread of the virus.  
In addition, we note that \covidRelated apps are not flagged by malware detectors. 
Yet, we found that some of them have been removed from \gp{}. 
Then, after assessing the complexity of \covidRelated apps and comparing them with
standard apps, we found that on average, \covidRelated apps are less complex,
which has been shown to often be indicative of quality for 
apps~\citep{jovst2013using}.
Finally, we applied state-of-the-art security and privacy scanners 
in order to check for potential cryptographic API misuses and data leaks in the code.

\subsection{Takeaways for different audiences}

\textbf{Broad public:} Our empirical study builds on a dataset collected strictly based on keywords, which we later curated to have an accurate but complete view on the development of apps to address \covid challenges. A main finding of our study is that our taxonomy of \covid apps highlights a variety of purposes, while the current definition of \covid apps on Wikipedia remains narrowly focused on “contact-tracing”. We expect that this systematic inference of knowledge from app markets will provide a broader understanding of what is a \covid app to the general public. More specifically yields, our study yields a characterization of \covidRelated apps for researchers, discusses qualitative analyses of \covidRelated apps for developers and provides useful information about \covidRelated apps exposed to end-users.

\textbf{[Researchers]} The \covid outbreak is a singular event in modern history that affected various domains. In the industry of software development, it is a rare occasion where a worldwide effort was put to produce apps to address the pandemic. Datasets and insights of this sudden and widespread production are valuable to the research community.  Our empirical study offers a first look at \covidRelated apps, which are developed in a fast-paced manner, for a variety of purposes, for different user groups, on-behalf of different stakeholders, with different business models, etc. We provide a taxonomy in this respect. 

\textbf{[Developers]} Development of apps is a continuous effort that can be guided with lessons learned from the successes and failures of similar apps. Our study yields a number of insights and empirical discussions on code quality and privacy issues. We even show that Google applies some strict policies to remove “undesirable” \covidRelated apps from its market, providing a clear warning of what developers should pay attention to when planning to develop \covidRelated apps. We further provide at the end of this study, different empirical results related to code quality and privacy. Typically, we discuss how developers can expose potential sensitive information that can be leaked during app execution.

\textbf{[End-users]} Our taxonomy, by highlighting the purposes of different \covidRelated apps, their processing of private information and the medium used to share information, provides a clear,  informative and multi-dimensional view for users wishing to adopt \covidRelated apps.

In summary, we present the following contributions:
\begin{itemize}
    \item We present the first systematic study of \covidRelated apps that
        explores their characteristics and compare them with other non-\covidRelated apps.
    \item We build a taxonomy of \covidRelated apps based on their described goals.
    \item We apply literature analysis tools on \covidRelated apps and discuss their results.
\end{itemize}

All artifacts are made available online at:
\begin{center}
\url{https://github.com/Trustworthy-Software/APKCOVID}    
\end{center}

\subsection{Research questions}
\label{es:research_questions}

The main objective of this work is to analyze and understand \covidRelated
apps.
To do so, we empirically observe apps characteristics by extracting features
that provide insights toward understanding those apps.
Hence, to accomplish this objective, we plan to answer the following research
questions:

\begin{itemize}
    \item \textbf{RQ 1:} What are \covidRelated apps used for?
   % \item \textbf{RQ 2:} When did \covidRelated apps start to appear in \gp{}? 
    \item \textbf{RQ 2:} Do \covidRelated apps have specific characteristics?
    \item \textbf{RQ 3:} Are \covidRelated Android apps more complex than standard apps?
    \item \textbf{RQ 4:} To what extent were \covidRelated apps removed from
        the official \gp{} and why?
    \item \textbf{RQ 5:} Who are \covidRelated apps' developers?
    \item \textbf{RQ 6:} Do \covidRelated apps have security issues?
    %Are \covidRelated apps flagged by privacy \& security scanners?
\end{itemize}

The remainder of this paper is organized as follows.  
First, we present an initial Android apps dataset and some
background for the reader in Section~\ref{background_dataset}.
Then, in Section~\ref{experiment_setup},
we give details about the experimental setup of our study.  
In Section~\ref{results}, we provide the results obtained from our experiments.
We discuss the threats to validity of our study in Section~\ref{discussion}.
Finally, we discuss related work in Section~\ref{related_work} and conclude in
Section~\ref{conclusion}.

%% file: background.tex
\section{Dataset and Background on App Analysis}
\label{background_dataset}

\textbf{Initial Dataset:} In order to perform our experiments and to exhaustively answer the research questions presented in Introduction, we need to rely on a comprehensive dataset.
Consequently, we used the state-of-the-art largest dataset available,
namely \az{}~\citep{Allix:2016:ACM:2901739.2903508}.
At the time of writing, \az{} contains more than 11 million Android apps 
(June 2020) that have been collected from different sources, such as \gp{} and 
other third-party providers (F-Droid, Anzhi, AppChina, etc.).
As it is continuously growing, researchers still heavily rely on \az{} for 
collecting apps and experimenting on it~\citep{ranganath2020free,shar2020experimental,he2020diversified,xu2020manis}.

\begin{figure}[t]
    \centering
    \captionsetup{justification=centering}
    \includegraphics[width=0.7\linewidth]{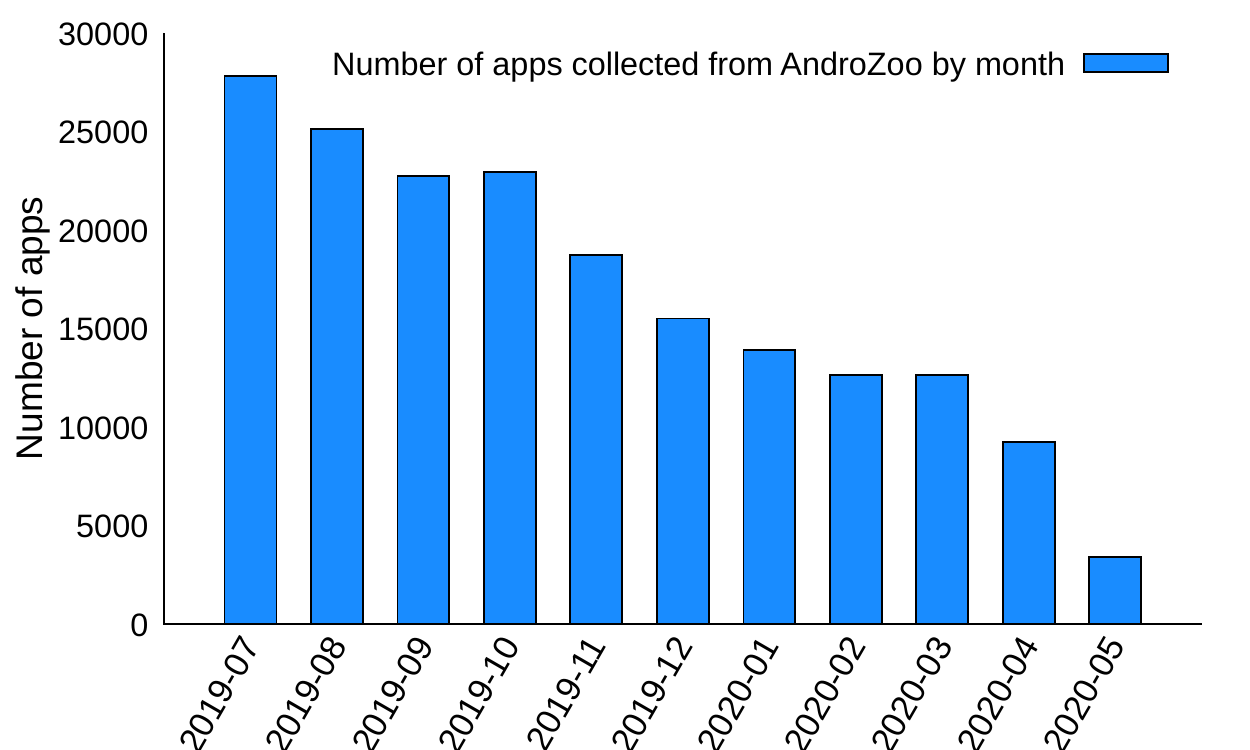}
    \caption{Monthly distribution of Android apps considered in this study}
    \label{fig:apps_dates_distribution}
\end{figure}

Since the \covid outbreak is quite recent, we did not consider the 11 million Android apps present in \az{}.
For our study, we instead considered and collected from \az{} all the apps 
ranging from July, 1\textsuperscript{st} 2019 to May, 25\textsuperscript{th} 
2020, leading to a total number of \appsdate{} collected apps.
Note that \az{} does not contain information about the release date of an app, 
i.e., there is no information about when an app has been uploaded on the market. 
For this reason, we approximate the date of an app by considering the date of 
the dex files in apks. 
Figure~\ref{fig:apps_dates_distribution} shows the distribution of collected 
APKs according to the month of the date of the dex files in APKs.

\noindent
\textbf{Dataset Augmentation:} 
It is a known problem that the date metadata in APKs (approximated as the 
date of the dex files) is not always reliable~\citep{moonlightbox}. 
For this reason, we decided to augment our dataset 
(1) by not considering the time window from July 2019 to May 2020 and 
(2) by selecting  any apps from \az{} whose package name contained
the words \emph{covid-19}, \emph{covid19} or \emph{coronavirus}.
With this heuristic, we were able to retrieve \appspkg{} apps from \az{}.

$\Rightarrow$ Finally, our initial dataset is constituted of \totalApps{} 
Android apps.

\noindent
\textbf{Short Background on Android Apps:} 
In general, Java is the language used by developers to create Android applications, 
though other languages can be used such as Kotlin, or even languages such as C 
and C++ thanks to the Native Development Kit (NDK).
For common Java applications, the source code is compiled into Java bytecode 
and run into the Java virtual machine.
However, Android apps are compiled into Dalvik bytecode (Dalvik Executable, i.e., Dex), which is the distribution format of app code in app packages. This bytecode is then targeted to be executed on the Android Android Runtime (which now replaces Dalvik Virtual Machine on recent Android versions).

Still, the DEX files do not constitute an application.
Indeed, an Android app is a collection of files packaged together in the so-called Android Application Package (APK).
This format is used for distributing apps on devices, not only via 
official and unofficial markets but also via any other channels where 
an independent file can be distributed.
Indeed, Android devices allow users to install applications from any 
sources once a configuration option is toggled off.

An APK is typically a zip file containing the following files:
(1) Metadata files,
(2) the certificate(s) used to sign the application,
(3) a lib folder containing platform-dependent compiled code,
(4) compiled and non-compiled resource files,
(5) one or multiple DEX files, generally named \texttt{classes$X$.dex} with $X$ an integer,
(6) an AndroidManifest.xml file describing the application (package name, components, version, access rights, etc.).

%% file: experiment_setup.tex
\section{Experiment Setup}
\label{experiment_setup}

In this section, we describe the setup of our experiments.
More specifically, in Section~\ref{subsec:curation} we describe how we create a dataset of \covidRelated apps by searching apps in both our initial dataset extracted from \az, and on the web. 
Then we describe in Section~\ref{es:features} how the information needed for 
answering the research questions was extracted.

\subsection{Dataset curation}
\label{subsec:curation}

The large majority of apps contained in our initial dataset of Android apps are not related to the \covid outbreak. 
Consequently, we need to curate this dataset. 
Our first idea was to rely on a clear definition of what is a \covidRelated app. 
However, we realized that finding a correct and precise definition is not obvious. 
For instance, the english Wikipedia page~\footnote{https://en.wikipedia.org/wiki/COVID-19\_apps} related to \covid ~\citep{wikipediacovid} defines \covidRelated apps as
\emph{"mobile software applications that use digital tracking to aid contact tracing in response to the COVID-19 pandemic, i.e. the process of identifying persons ("contacts") who may have been in contact with an infected individual."}.
We quickly considered this definition as too restrictive since we found several \covidRelated apps that are not about "digital contact tracing". 
As a result, rather than relying on a definition of what is a \covidRelated app 
we (1) implemented several heuristics based on assumptions, and 
(2) performed several quality checks  to filter out irrelevant apps. 

\textbf{Assumption 1:} Our first assumption is that a \covidRelated app contains 
\emph{strings} (e.g., in a class name or method name) related to \covid. 
Based on this assumption, we defined several regular expressions that we apply 
on various fields on an Android app. 
More specifically,  we proceed as follows:\\
Let $A$ be a set of Android apps and $apk$ an app of this set. Let be:
\begin{itemize}
	\item[\textbullet] $C_{apk}$ the set of classes names in $apk$,
	\item[\textbullet] $M_{apk}$ the set of method names in $apk$,
	\item[\textbullet] $F_{apk}$ the set of file names in $apk$,
	\item[\textbullet] $S_{apk}$ the set of strings contained $apk$,
\end{itemize}
We keep $apk$ if at least one element of $C_{apk} \cup M_{apk} \cup F_{apk}\cup S_{apk}$
matches at least one regular expressions listed in 
Table~\ref{table:regularExpressions}.  
An app is kept if for instance, a class name contains the substring $coronavirus$, or if the app contains a string with $pandemi$ as substring. 
\begin{table}[ht]
    \centering
    \begin{tabular}{l|l}
"(?i).*coronavirus" & "(?i).*corona" \\
"(?i).*sars(-?cov)" & "(?i).*quarantin.*" \\
"(?i).*lock-?down" & "(?i).*containment" \\
"(?i).*social-?distanc.*" & "(?i).*pandemi.*" \\
"(?i).*out-?break" & "(?i).*epidemi.*" \\
"(?i).*confinement" &
    \end{tabular}
    \caption{Regular expressions used to filter out non \covidRelated apps}
    \label{table:regularExpressions}
\end{table}

\begin{figure}[h]
    \centering
    \captionsetup{justification=centering}
    \includegraphics[width=\linewidth]{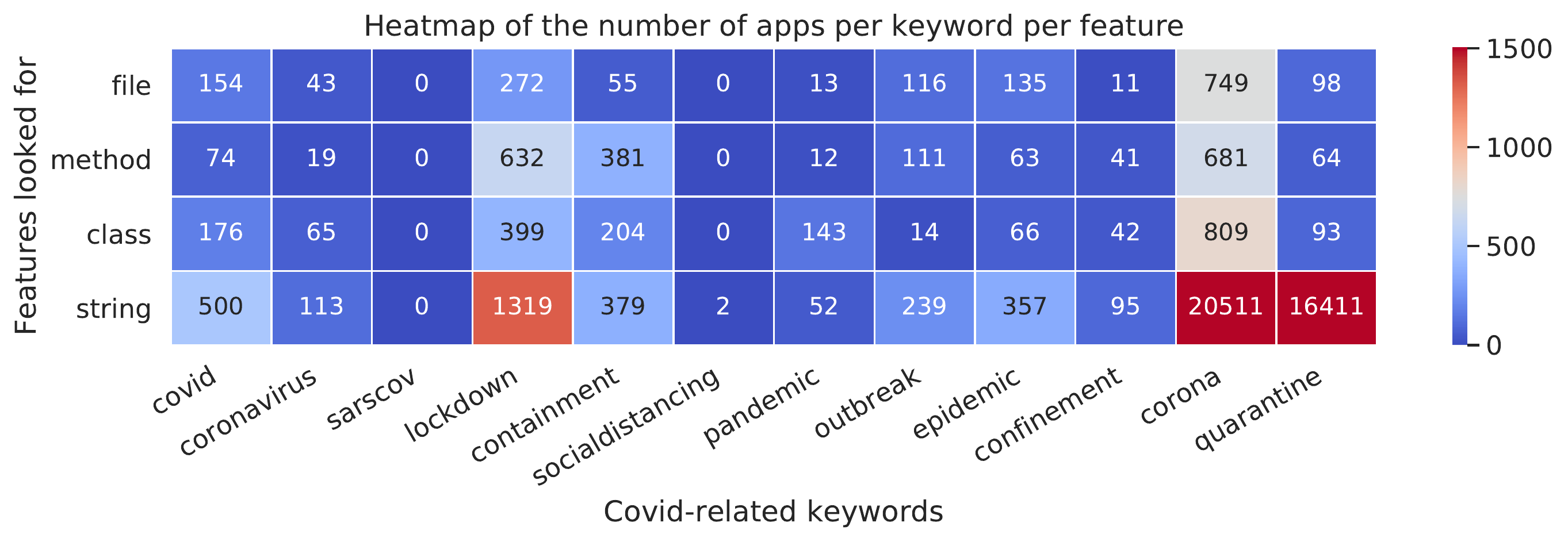}
    \vspace{-3mm}
    \caption{Heatmap representing the number of apps gathered by each 
    \covidRelated keyword, and for each family of features}
    \label{fig:heatmap}
\end{figure}

We ended up with a set of \numCovidApps{} supposedly \covidRelated 
apps.
In Figure~\ref{fig:heatmap}, we can see the number of apps that were 
retrieved per keyword (note that several keywords can be present in a given app).

\textbf{Quality Check \#1:}
We can see in Figure~\ref{fig:heatmap} that the keywords \emph{corona} and \emph{quarantine}
match a significantly larger number of apps.
After investigation, we found that the keyword \emph{quarantine} is used
in \texttt{strings} 
by developers to check if the smartphones are \emph{rooted}, i.e., 
if the user has super-user (\emph{root}) privilege. 
Indeed, the presence of some packages can characterize the fact that a 
smartphone is rooted. 
Among such packages, one is named \texttt{com.ramdroid.app\-quarantine} and 
another is named
\texttt{com.ramdroid.appquaran\-tinepro}.
To ve\-rify the presence of these two packages, developers often use these package 
names as string, 
and thus the regular expression using \emph{quarantine} matches.
This explains why there is such a high number of apps containing \emph{quarantine} in strings.
We manually analyzed several dozens of apps and confirmed they are not real \covidRelated apps (mainly music, education, news, shopping, and game apps).

Regarding the \emph{corona} keyword, it also appears in a substantial number of apps, especially in string feature.
After manual investigation, we found that apps retrieved with this keyword 
use a framework called \emph{Corona} developed by 
Coronalabs\footnote{https://coronalabs.com/}. 
These apps are mainly games, entertainment 
and personalization apps.
Moreover, the word \emph{corona} not only refers to \covid, but also has uses in 
several unrelated contexts (in architecture, beverages, books, music, movies, and games).

We decided to rule out both keywords \emph{corona} and \emph{quarantine},
 as they revealed to mostly bring noise in our dataset. 
After filtering the apps gathered, i.e., not taking into account the 
\emph{corona} and \emph{quarantine} keywords, we obtained \appsAfterFilter{} apps.

\vspace{2mm}
\textbf{Assumption 2:}
Our second assumption is that since \az is known to contain successive 
versions of the same apps~\citep{Allix:2016:ACM:2901739.2903508,moonlightbox}, 
our dataset contains successive versions of the same \covidRelated app.
Therefore, for the subsequent analyses, we only keep the latest version of a given app.
This step is performed by comparing apps version code---available in \az{} 
metadata---and keeping the highest value.
After this step, our dataset contains \appsAfterUniqVersion{} apps.

\begin{figure}[ht]
    \centering
    \captionsetup{justification=centering}
    \includegraphics[width=0.6\linewidth]{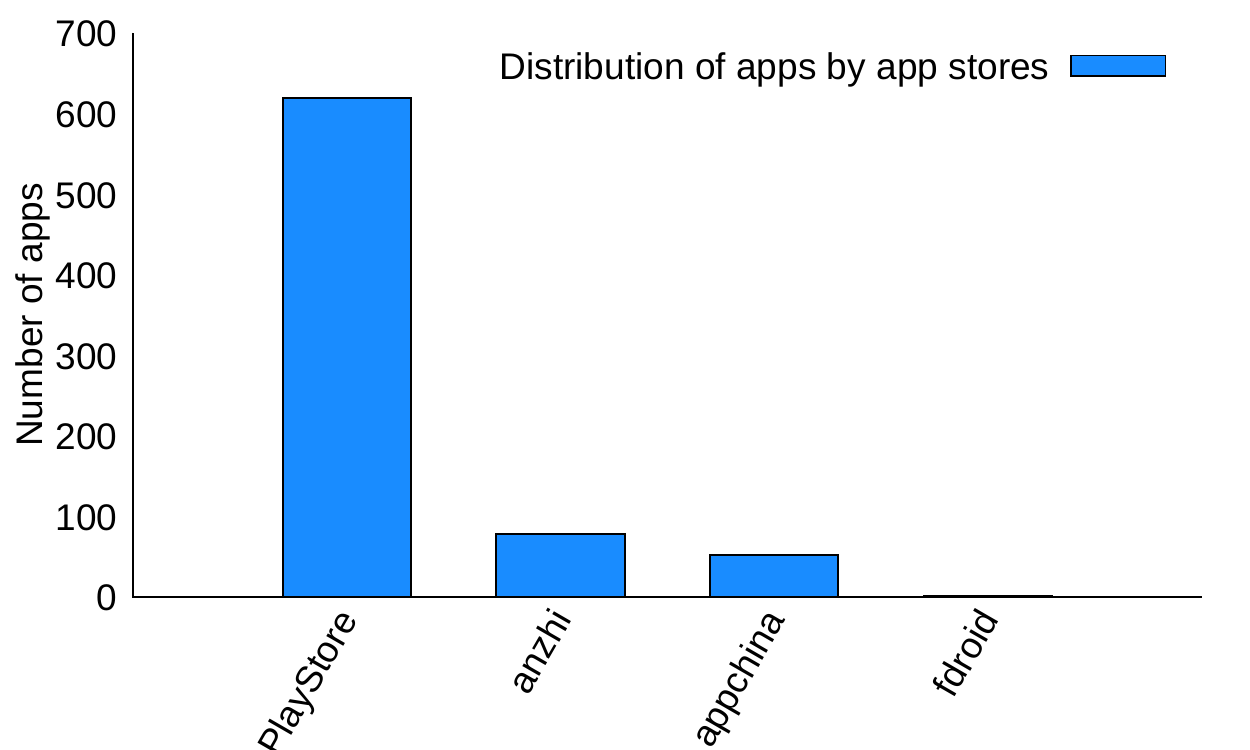}
    \caption{Distribution of the \appsAfterUniqVersion{} supposedly \covidRelated apps by markets where they were obtained from}
    \label{fig:stores}
\end{figure}

\vspace{2mm}
\textbf{Assumption 3:}
Our third assumption is that \emph{official} \covidRelated apps are released 
only on the official Android market, i.e., the \gp{} market.
Figure~\ref{fig:stores} depicts the distributions of the apps per market from where they have been crawled. 
Most of the apps were released in the official \gp{} market.
By considering apps from \gp{} only, we reach the number of 619 apps.

\begin{figure}[ht]
    \centering
    \captionsetup{justification=centering}
    \includegraphics[width=0.7\linewidth]{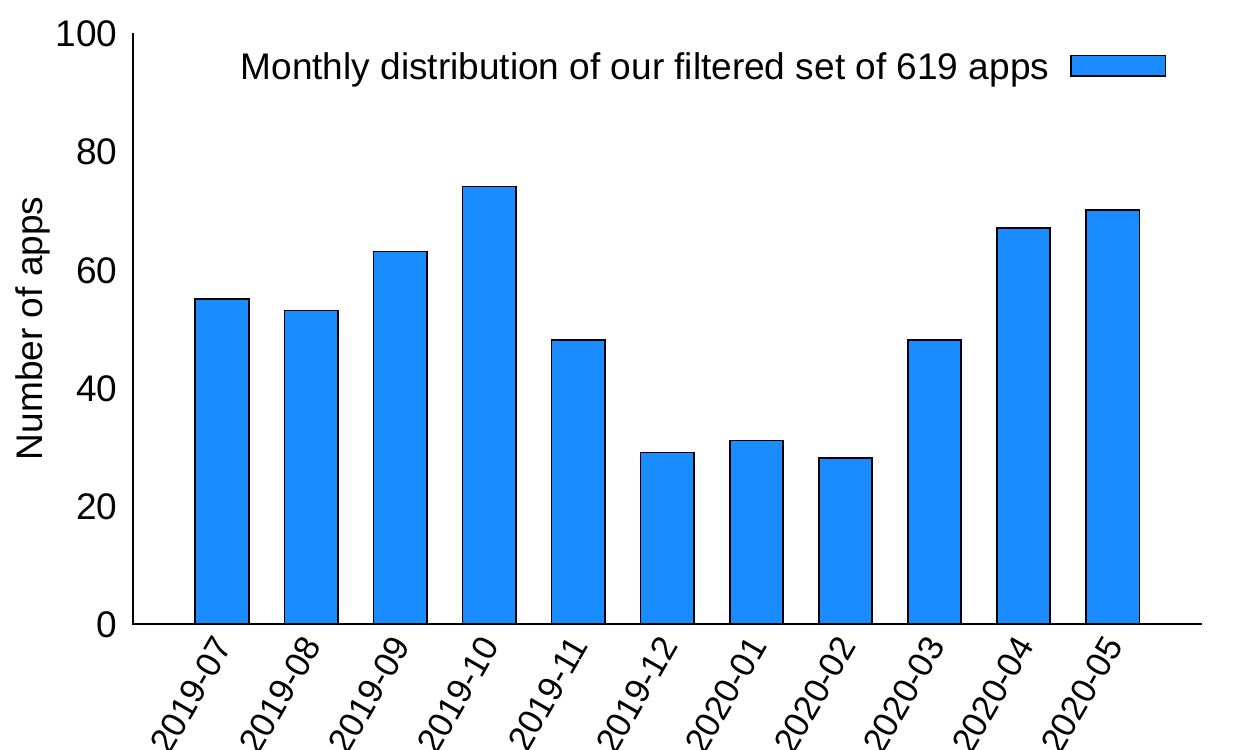}
    \caption{Timeline representing the number of apps with \covidRelated keywords in function of time}
    \label{fig:timeline_covidrelatedapps}
\end{figure}

\vspace{2mm}
\textbf{Quality Check \#2:}
To check the quality of our dataset, we inspect the date of the dex file in the APKs. 
The \covid outbreak was not known until the very end of 2019, so we expect 
to find apps from 2020 only. 
Figure~\ref{fig:timeline_covidrelatedapps} presents the distribution of 
those apps by the date of the dex file in the APK.
First, we observe that fewer apps were released during winter 2019-2020.
Then we can see that starting from March 2020, a significant amount
of apps were developed, which corresponds to the containment period.

Finally, Figure~\ref{fig:timeline_covidrelatedapps} shows that our set of \covidRelated apps contains a 
significant amount of apps developed before the pandemic, which indicates 
the low quality of our dataset and  
that an additional layer of filtering is needed to only keep \covidRelated apps.
Indeed, our manual investigations of those apps revealed that they contained 
keywords that may be overly broad. 
For example, it appeared that keywords such as \emph{outbreak}, 
\emph{pandemic}, \emph{containment} or \emph{lockdown} were 
popular in games, most notably in the zombies and survival games genres.

\vspace{2mm}
\textbf{Assumption 4:}
Our fourth assumption is that the description of an app is a reliable source 
of information to check that an app is \covidRelated.
\az does not provide the app descriptions, 
so we queried \gp{} to retrieve the description of the 619 apps. 
Note that it was possible to get the description for 537 apps, while 
82 apps were not available anymore in \gp{} at that time. 
Actually, \az uses a crawler to automatically download apps. 
It is possible that \az downloads an app at a time $t_1$, but if 
Google Play decides to  remove this app  (or if the developer decides to 
remove it) at time $t_2$, 
with  $t_1<t_2$, it is not possible to access the description of the 
app anymore after time $t_2$.
We found that \ggl{} is actively deleting apps that are violating their policy with respect to  \covid~\citep{googlecovid, googleupdate}.

Manual investigations of the remaining 537 apps were conducted to qualify an app
as \emph{\covidRelated}.
We did so by analyzing the \gp{} page of every app, reading the description,
and looking at screenshots.
We did not encounter any ambiguous case, hence it was straightforward
to qualify an app as \covidRelated or not.
With this method, we determined with high certainty that \covidAppsInAz{} 
apps taken from \az{} were \covidRelated.

Note that the set of apps from our initial dataset for which the package 
name contained \covidRelated keywords contained \appspkg{} elements and we
only retrieved \covidAppsInAz{} apps.
This is explained by the fact that among those \appspkg{} apps, there 
were, for almost one-fifth, different versions of the same app.
In addition, we were not able to analyze the \gp{} page
of apps that were not available in the \gp{} market.

\begin{figure}[ht]
    \centering
    \includegraphics[width=.49\textwidth]{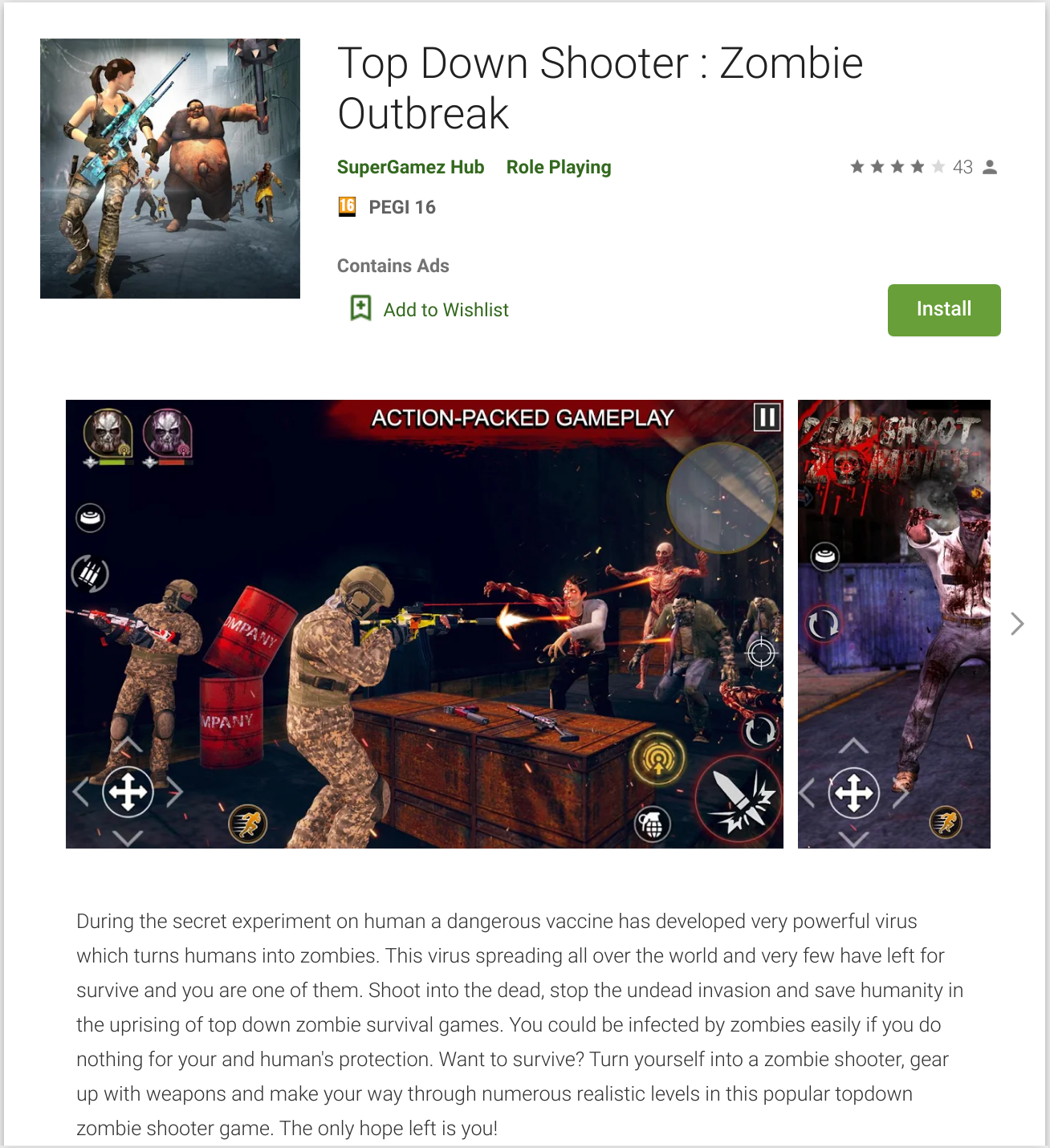}
    \includegraphics[width=.49\textwidth]{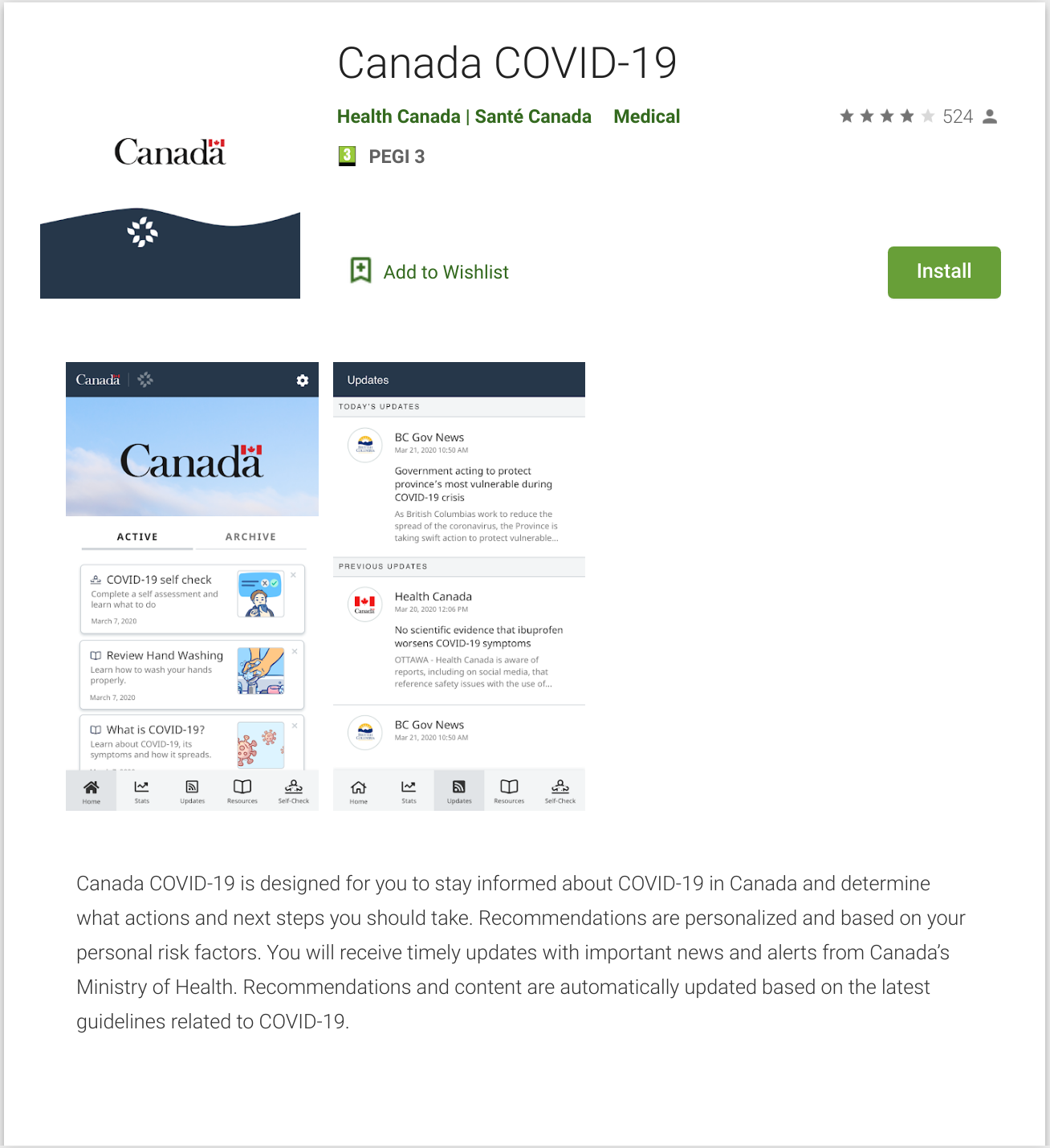}
    \caption{Comparison between two \gp{} page of two different apps. Left app: Zombie game; right app: \covidRelated app.}
    \label{fig:comparison_apps}
\end{figure}

Figure~\ref{fig:comparison_apps} gives two examples to show an app we discarded
and an app that was considered as \covidRelated.
First, the game picture on the left clearly shows a game in which the user
has to kill zombies, even though the game contains the string \emph{outbreak}.
Second, the \covidRelated app on the right is explicit regarding the content
delivered to the user. The title refers to \covid, the description gives
clues about the content, i.e., information about \covid and guidelines about
\covid.
Besides, the screenshots are also explicit by depicting what actions 
users can perform, here \covidRelated actions, i.e., getting information about
\covid, performing self-diagnosis, receiving guidelines, and news about \covid.
It represents how unequivocally our decisions were to be made
to qualify an app as \covidRelated or not. 

\textbf{Assumption 5:}
As \az{} is not exhaustive and all the apps in the \gp{} market are not
available for download from every country (\az{} crawls from specific countries),
we felt the need to expand our research.
Consequently, manual investigations were conducted on the web to search 
for apps that would not be 
available in Google Play and/or that our empirical analysis on AndroZoo did not catch.
We found \additionalApps{} additional apps from diverse countries from
different time span, e.g. we found apps that were first seen in June 2020.

We verified if any of those 48 additional \covidRelated apps were already in Androzoo and might have been missed by our diverse filters, it was not the case for all of them.
Thus, our set of COVID-applications reached a length of \covidRelated apps, each one of them from \gp{}.

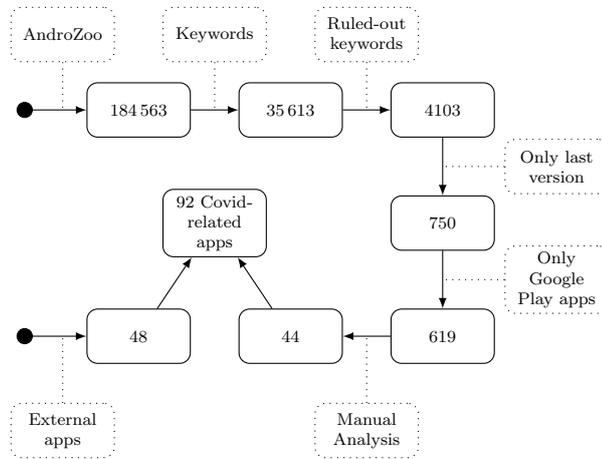
\begin{figure}[h]
    \centering
    \input{dataset}
    \caption{Process of our dataset curation. Numbers represent numbers of apps. Dotted boxes represent filters used to refine the dataset.}
    \label{fig:dataset}
\end{figure}

$\Rightarrow$ Figure~\ref{fig:dataset} summarizes our dataset curation process.

\subsection{When did \covidRelated apps start to appear in \gp{}?}

\label{res:dates}

\begin{figure}[ht]
    \centering
    \captionsetup{justification=centering}
    \includegraphics[width=.7\linewidth]{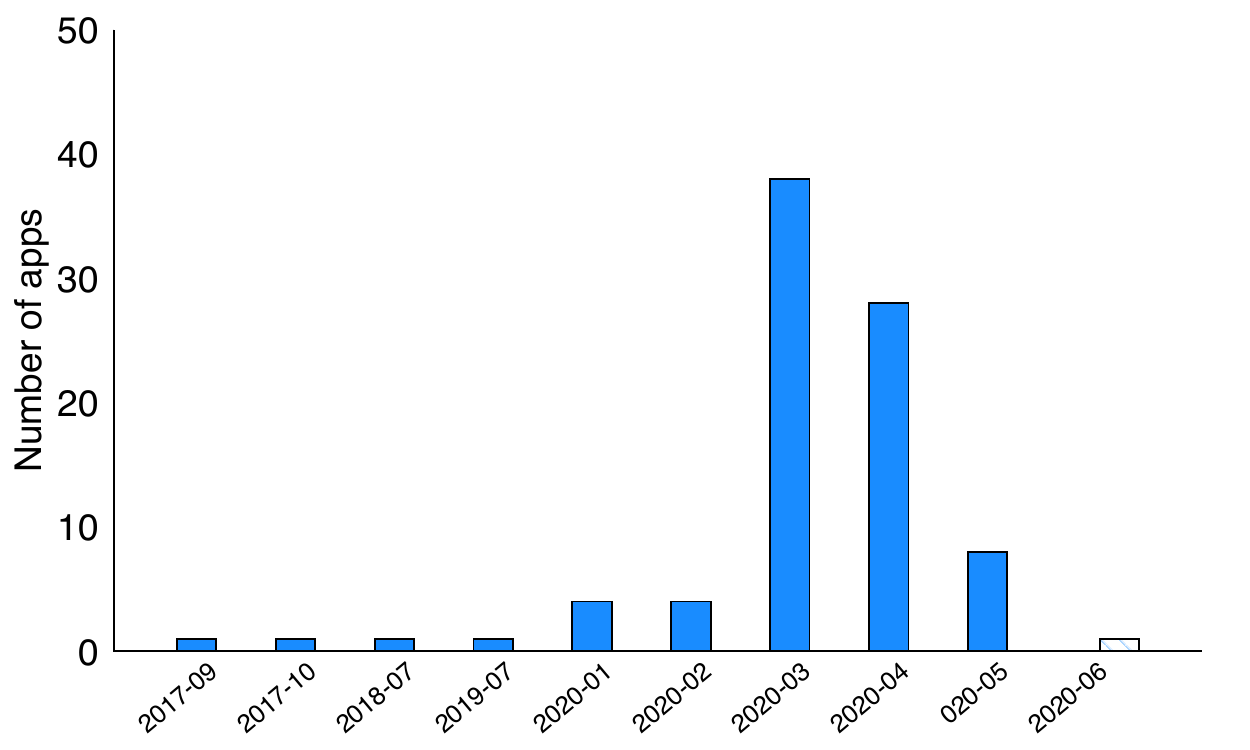}
    \caption{Number of \covidRelated apps by appearance date. The month of June appears hatched to represent the fact that our study was conducted in early June, which means we do not have all apps released in June.}
    \label{fig:timeline_covidapps}
\end{figure}

In this section, we consider the 92 collected \covidRelated apks. As the first release date of apps is not available on \gp{}, we need to rely on other sources of information to try to find the first appearance date of a given app. Therefore, in order to visualize when \covidRelated apps appeared, we considered, for each app (when available), the earliest date among the AndroZoo added date, the first seen dates in VirusTotal and in third party datasets (i.e., Koodous\footnote{\url{https://koodous.com}}, APKCombo\footnote{\url{https://apkcombo.com}}, APKPure\footnote{\url{https://apkpure.com}}, and AppBrain\footnote{\url{https://appbrain.com}}).The results can be seen in Figure\ref{fig:timeline_covidapps}.

First we can see that four apps were released long before 2020 (com.intelli\-gent.alertaguate, br.gov.datasus.guardioes, pl.nask.mobywatel and co.gov.ins\-.guardianes), i.e., two in 2017, one in 2018 and one in 2019. After investigation, we found that these apps were originally developed as governmental health apps and were updated to account for the pandemic specificities.

In 2020, we can see that a few \covidRelated apps started to appear as early as January 2020, i.e., before the pandemic was officially recognized by the World Health Organization (i.e., March, 11\textsuperscript{th} 2020) ~\citep{whopandemic}.

We also note from Figure ~\ref{fig:timeline_covidapps} that in March 2020, the number of new apps increased drastically.
Finally, since we stopped the collection process in early June 2020 (prior to submission), only a few \covidRelated apps have been collected this month. Our artefact of collected apps however has been updated with new apps that appeared after our submission. Evidently, these apps could not be taken into account in our empirical analysis.

\subsection{Features extraction}
\label{es:features}

In this section, we expose what features were extracted from the apps considered in this study and how we extracted it.
Three ways of retrieving the needed information were considered: 
(1) Extracted from \az{}, 
(2) Automated analysis of the apps, and 
(3) Manual analysis, which we describe below.

\subsubsection{\az{} metadata}

Apps from \az{} are provided with additional metadata, i.e., a vector of length 11 representing different information.
Those metadata elements are used to expose several properties of the apps under analysis.
We consider the date of the dex files to verify when the \covidRelated apps started to appear.
The package names and version codes were useful to have some insights into the versioning of the apps and to keep only the latest version of the app considered, i.e., one version of the app will be analyzed.
When available, the number of AntiVirus products reporting an app as malicious 
(obtained from \vt{}) was used for qualifying the maliciousness of the apps.
Finally, \az{} metadata also indicate the source where an app was obtained from.

\subsubsection{Automated analyses of apps}

\az{} metadata being limited, we additionally leveraged existing tools 
and frameworks to analyze Android apps in order to obtain the information 
to acquire the information needed by our study.

\begin{itemize}
	\item We developed a program relying on the \ag{} software package~\footnote{https://github.com/androguard/androguard}. 
	Thanks to this tool, we extract the permissions requested by the apps, as well as information about components in the applications (i.e., Activities, Services, etc.).
	We also automatically compute the complexity metrics described in Appendix~\ref{appendix:complexity}.
	\item We used CogniCrypt~\citep{kruger2017cognicrypt} and its headless implementation CryptoAnalysis~\citep{cryptoanalysis} to check that they were written following best practices regarding cryptographic APIs.
    \item As several \covidRelated apps seek to acquire and process 
    the location of users (considered a piece of sensitive information), 
    we verify if those apps were subject to data leaks. 
    Hence, we leverage FlowDroid-IccTA~\citep{li2015iccta}, the 
    state-of-the-art static data leak detector dedicated to Android apps.

\end{itemize}

\subsubsection{Manual analyses of apps}
\label{manual_analyses}

While application information that can be readily obtained through 
automatic tools were necessary for our study, we went a step further 
and acquired \emph{qualitative} data on the apps. 

By collecting and carefully reading the descriptions of apps\footnote{The description of an app can be retrieved from the web page of the app on Google Play. This web page can be accessed with the following link: https://play.google.com/store/apps/details?id=PACKAGENAME, where PACKAGENAME is the package name of the app (e.g. \emph{fr.gouv.android.stopcovid}).} 
and 
by confirming the validity of our understanding by matching the reviews 
and app screenshots with the description, we were able to assemble 
highly-qualified data on the goals of app developers, on whether 
an app is developed by a state body or an individual person, etc.

Finally, after leveraging the automated tools described above, 
their results were manually confirmed for the set of 
\covidRelated apps (i.e., \covidrelatedApps{}). 
This allowed us to ensure those tools did not yield false-positives, 
and that their results were sound and consistent.

In the following, we give further details on how the manual analyses were performed:
\begin{itemize}
    \item App descriptions: The first author read each app description systematically extracting relevant information on the WHO, WHAT, WHEN and HOW that characterize the apps. To ensure that the collected information is reliable, the extraction is repeated for every app. Then he proposed a first summary based on a careful analysis of recurrence in characteristics. The author team subsequently convened to refine this categorisation and validate the final taxonomy.
    \item App properties: In order to verify some properties on apps, the first author decompiled each app (using JADX~\footnote{https://github.com/skylot/jadx}) and searched for the property to verify its use (e.g., is a certain keyword used in the context of a game or is the app really providing information or service about to Covid?)
    \item Tool results: The first author ran the third-party tools multiple times on \covidRelated apps. For instance, every output of \iccta{} was manually checked to confirm that it is not a false positive.
\end{itemize}

%% file: dataset.tex
\begin{tikzpicture}
    \tikzstyle{arrowStyle}=[-latex]
    \tikzset{node/.style={minimum height=0.9cm, rounded corners, text width=1.5cm,align=center,scale=.8}}

    \draw[fill=black] (-1.5,0) circle (0.1);
    \draw[fill=black] (-1.5,-3) circle (0.1);
    \path
    (0,0) node[node,draw] (184563) {\totalApps{}}
    ++(2,0) node[node,draw] (35613) {\numCovidApps{}}
    ++(2,0) node[node,draw] (4103) {\appsAfterFilter{}}
    ++(0,-1.5) node[node,draw] (750) {\appsAfterUniqVersion{}}
    ++(0,-1.5) node[node,draw] (619) {619}
    ++(-2,0) node[node,draw] (44) {\covidAppsInAz{}}
    ++(-2,0) node[node,draw] (48) {\additionalApps{}}
    ++(-1,4) node[node,draw,dotted] (androzoo) {\az{}}
    ++(2,0) node[node,draw,dotted] (keywords) {Keywords}
    ++(2,0) node[node,draw,dotted] (ruledout) {Ruled-out keywords}
    ++(2.5,-1.75) node[node,draw,dotted] (version) {Only last version}
    ++(0,-1.5) node[node,draw,dotted] (googleplay) {Only \gp{} apps}
    ++(-2.5,-2) node[node,draw,dotted] (manual) {Manual Analysis}
    ++(-4,0) node[node,draw,dotted] (external) {External apps}
    ++(2,2.75) node[node,draw] (covidrelated) {\covidrelatedApps{} \covidRelated apps};

    \draw[->,>=latex] (-1.5,0) -- (184563);
    \draw[->,>=latex] (-1.5,-3) -- (48);
    \draw[->,>=latex] (184563) -- (35613);
    \draw[->,>=latex] (35613) -- (4103);
    \draw[->,>=latex] (4103) -- (750);
    \draw[->,>=latex] (750) -- (619);
    \draw[->,>=latex] (619) -- (44);
    \draw[->,>=latex] (44) -- (covidrelated);
    \draw[->,>=latex] (48) -- (covidrelated);
    \draw[dotted] (androzoo) -- (-1,0);
    \draw[dotted] (keywords) -- (1,0);
    \draw[dotted] (ruledout) -- (3,0);
    \draw[dotted] (version) -- (4,-0.75);
    \draw[dotted] (googleplay) -- (4,-2.25);
    \draw[dotted] (manual) -- (3,-3);
    \draw[dotted] (external) -- (-1,-3);

\end{tikzpicture}

%% file: results.tex
\section{Results}
\label{results}

In this section, we present our experimental results and we answer 
our research questions.

\subsection{What are \covidRelated apps used for?}
\label{covid_used_for}

\textbf{Motivation.}
The sudden increase of \covidRelated apps during the pandemic shows that Android apps developers have been active in providing end-users with solutions to address the \covid pandemic.
Nevertheless, the functionalities of \covidRelated apps are not known and have not been studied.
In this section, we study, characterize, and build categories from which \covidRelated apps belong.
The categorization of \covidRelated apps offers a first layer of knowledge toward understanding them.
The outcome of this research question will give an overview of \covidRelated apps' functionalities to the general public.

\noindent
\textbf{Strategy.}
Textual descriptions of apps on markets generally provide a wealth of
information on the purpose and functionalities that developers advertise.
We undertake to systematically examine the descriptions of all the apps under study. Unfortunately, since \gp{} is actively moderating \covidRelated
apps, we have faced an issue with some apps that we were able to initially
collect but which were no longer available on the market at the time of
analysis. 
Eventually, our analysis of descriptions was performed on \covidrelatedAppsWithDesc{}
apps. 
In other words, from the time we read the descriptions of the apps to curate
our dataset (as explained in Section~\ref{subsec:curation}) and the time we
perform this more in-depth study, i.e., collecting information related to
the features of the apps, 14 apps ($92-78$)  were not able anymore on
GooglePlay.

\noindent{\bf A Taxonomy of \covidRelated apps.} After a careful analysis of
information available in \gp{}, we summarize for each app its general goal,
i.e., which aspects of the \covid crisis the app is precisely intended to
address.
Eventually, we identified three main categories to which each app can
be associated with possible overlap between categories, i.e., an app can
be associated with several categories:

\begin{enumerate}
    \item {\em Information broadcast (top-down)} - Apps in this category aim to provide users with various types of information, from general guidelines, infection statistics to general \covid news. Although such apps are not always officially released by government bodies, they often relay official information from top (authorities) down (users).
    \item {\em Upstream collection (bottom-up)} - Apps in this category collect information from users and make it available to the developer and/or an official body, such as a country's health authorities.
    \item {\em Tooling} - Apps in this category serve as tools with functionalities that directly deal with daily aspects of the \covid (e.g., generation of certificates).
\end{enumerate}

\noindent{\bf [1] Information broadcast.}
From the collected dataset of \covidRelated apps we identified several distinguishing scenarios in apps performing information broadcasting. Figure~\ref{fig:info} overviews
the related characteristics, notably based on the types of information that are made
available to the user: 

\begin{figure}[!h]
    \centering
    \input{info}
    \caption{Information Broadcast category. The number below a leaf box indicates the number of related apps.}
    \label{fig:info}
\end{figure}
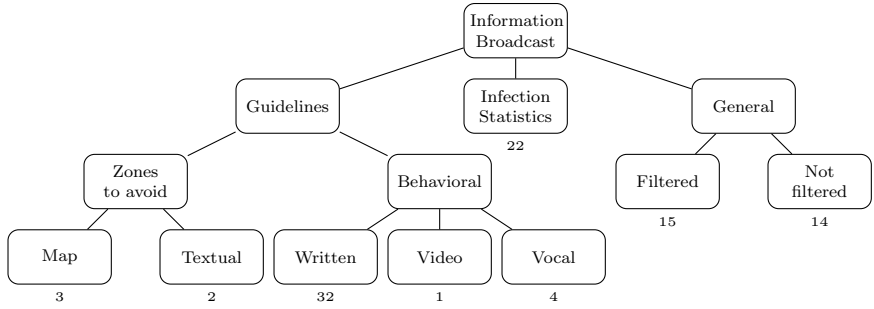

\begin{enumerate}
    \item {\em Guidelines} on measures to take to minimize the risk of infection - Among such apps, some render maps highlighting high-risk areas. Other apps provide behavioral advice (e.g., how to wash hands), leveraging the whole spectrum of available media: (1) textual descriptions (for the majority of apps), (2) videos and, (3) audio clips. 
    \item Continuously-updated {\em Statistics} on the pandemic evolution;
    \item {\em General information} about \covid, such as about the typical symptoms. We identified two different scenarios in the provision of general information:
    \begin{itemize}
    	\item Some apps present {\em curated information}, i.e., information that is somehow checked and filtered by the development team before it is shown to the public. Such information is often tagged in a way that allows interested people to find the source, and gauge its credibility. Sometimes, these apps are developed directly by an entity that itself carries credibility as a source of information, such as national healthcare authorities.
            An example is the \emph{MyHealth Sri Lanka} app\footnote{package name: \emph{app.ceylon.selftrackingapp}} developed by the national ICT Agency, which 
            presents to the user \emph{verified information on 
            the current \covid status.}
    \item A number of apps appear to provide unfiltered information regarding \covid. Their developers are not always themselves entities that would traditionally be assumed to have any specific credibility on the matter. For example, the DiagnoseMe app\footnote{package name: \emph{bf.diagnoseme.fasocivic}}, which claims to provide the user with \emph{all the information on the virus}, is proposed by an association with unrecorded expertise in health. 
    \end{itemize}
\end{enumerate}

\noindent{\bf [2] Upstream collection.} Most apps in our dataset perform data collection from users. This suggests that many app providers consider \emph{data} to be key in the mitigation of the \covid crisis. App providers indeed collect a variety of information, including user personal information (e.g., name, age, address, etc.), 
some medical information (e.g., whether a user is infected with \covid, the  therapies that are used). Some apps are even used to keep a health diary (sharing information about symptoms every day), or to report the infection of people in the app user's acquaintances.

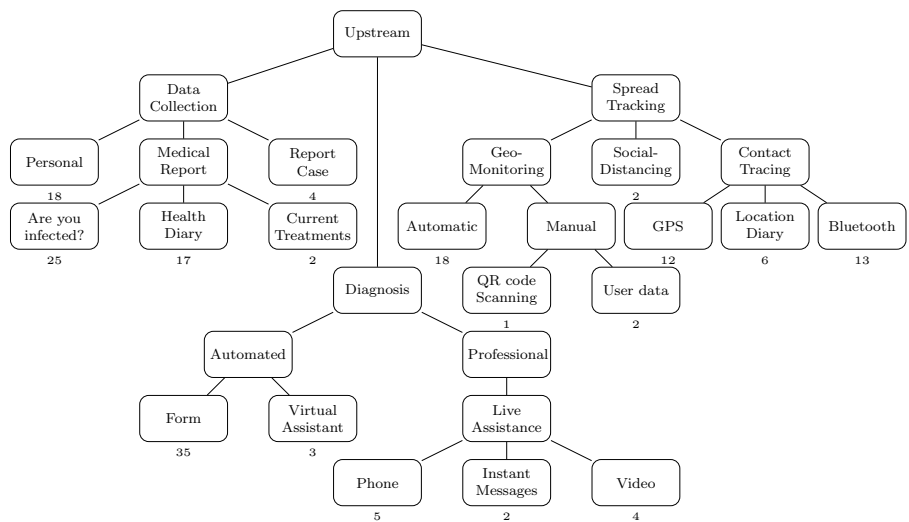
\begin{figure}[!h]
    \centering
    \begin{adjustbox}{width=\linewidth}
        \input{upstream}
    \end{adjustbox}
    %\vspace{-1mm}
    \caption{Upstream category}
    \label{fig:upstream}
\end{figure}

Overall,  we have identified three different ways in which apps collect user data, as summarized in Figure~\ref{fig:upstream}.
Note that in the case of {\em data collection} and {\em spread tracking} apps, we 
did not try to qualify whether apps were as privacy-preserving as their 
developers claimed they were (e.g., data is deleted after $N$ days), 
nor to determine to what extent the collected data is shared with 
third parties.

Similarly, for this paper, we did not analyze the inner workings of contact-tracing 
apps, and we did not evaluate the merit nor the opportunity of contact-tracing, 
this having already been---and still being to this date---discussed by 
security researchers~\citep{chris_culnane,anderson,baumgrtner2020mind}.

Several apps take inputs from the users to offer diagnoses related to \covid. Such apps can provide a built-in 
questionnaire that users have to fill within the app, or leverage a virtual 
assistant or chatbot. In these cases, the diagnosis can be made automatically, with no interaction nor confirmation 
with a trained medical practitioner. 

Other apps, however, provide a somewhat more traditional medical visit experience, 
by offering the facilities needed to remotely exchange (e.g., via instant text messages as well as voice and/or video calls) with a medical doctor. Such apps are used from home,  since millions of people worldwide were confined, and were potentially reluctant 
or unable to visit a brick-and-mortar doctor's office.

Additionally, some apps are developed to track the spread of the virus by locating the users of the apps.
While a few of those apps use simple geo-monitoring with GPS information for tracking users,  most  apps do it automatically. Nevertheless, we found a few apps that request users to provide a-posteriori
the locations they have visited on a given day. We also identified 
one app  which uses QR code scanning at the entrance of public buildings to obtain 
precise location information, while still being fully under users' control.

With respect to tracing, a few apps promote social-distancing using the GPS location of users, 
the goal being to not approach other people too closely.

Furthermore, several apps implement contract-tracing, i.e., the ability 
to retrieve who a specific person has been in contact with, providing 
users a way to know if they have encountered someone infected, and potentially infectious. 
Contact-tracing apps mainly rely on three methods, (1) Using the GPS location of users, 
(2) Using the Bluetooth technology to detect proximity, and 
(3) Using a location diary that the users have to manually fill.

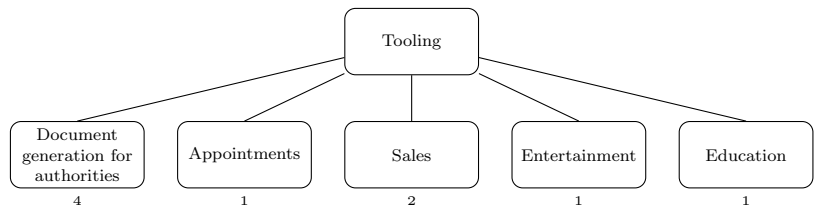
\begin{figure}[ht]
    \centering
    \input{tooling}
    \caption{Tooling category}
    \label{fig:tools}
\end{figure}

\vspace{5mm}
\noindent{\bf [3] Tooling}
The last category is the \emph{tooling} category which includes several types of 
tools aimed at helping users deal with some consequences of the \covid crisis (see Figure~\ref{fig:tools}).
A few apps allow users to auto-generate documents for their local authorities 
(e.g., travel authorization that had been made mandatory in several countries during containment).

Users can also install apps offering appointment-capabilities for medical purposes, 
or selling \covidRelated products (e.g., masks, hand-sanitizers, etc.).  

On the entertainment front, apps were released proposing games around the 
pandemic, or providing users with \covid-themed image filters, for example adding 
a virtual mask, or adding virtual decorative elements to an actual mask.

Lastly, apps were also made to cater to the newly-discovered needs of 
massive remote education.

The interested reader can inspect Tables~\ref{first_part_table},~\ref{second_part_table} and~\ref{third_part_table} for more information about the 
mapping between categories and the apps for which we were able to retrieve the relevant information.

Table ~\ref{first_part_table} gives the list of the \covidrelatedAppsWithDesc{} \covidRelated apps from which the Google Play page existed and that we gathered during this study and the data we were able to extract from them. The second column gives the country of origin of each app, the third column gives information about the type of developer of each app (e.g., governmental, researcher, company, etc.) and the fourth column gives the target of each app (e.g., citizens,journalists, etc.). Afterwards, the rest of the table is composed of the first category of our taxonomy. We can see that each column of this category is a leaf node of the Information Broadcast branch of the taxonomy (see Fig.~\ref{fig:info}). A check mark indicates that the corresponding app belongs to this category. Table~\ref{second_part_table} also represents the list of \covidrelatedAppsWithDesc{} \covidRelated apps, but it shows the mapping between each app and the upstream branch (see Fig.~\ref{fig:upstream}). Finally, Table~\ref{third_part_table} lists all the \covidrelatedAppsWithDesc{} \covidRelated apps with the last part of the leaf node of the upstream branch and the tooling branch (see Fig.~\ref{fig:tools}).

\highlight{
    \textbf{RQ1 Answer:} Our empirical analyses show that overall, \covidRelated apps 
    are mainly developed for:
    \begin{enumerate}
        \item Providing users with information;
        \item Collecting data from users;
        \item Offering users \covidRelated tools.
    \end{enumerate}
    The characterization of \covidRelated apps from our empirical study offers a first layer of knowledge about \covidRelated apps. We encourage future research to dig deeper into knowing \covidRelated apps.
    Tables~\ref{first_part_table},~\ref{second_part_table} and~\ref{third_part_table} offer end-users an overview of each \covidRelated apps.
}

\begin{table}
    \input{first_part_characteristics}
    \caption{First part of \covidRelated apps' characteristics retrieved from \gp{} apps pages}
    \label{first_part_table}
\end{table}

\begin{table}
    \input{second_part_characteristics}
    \caption{Second part of \covidRelated apps' characteristics retrieved from \gp{} apps pages}
    \label{second_part_table}
\end{table}

\begin{table}
    \input{third_part_characteristics}
    \caption{Third part of \covidRelated apps' characteristics retrieved from \gp{} apps pages}
    \label{third_part_table}
\end{table}

\subsection{Do \covidRelated apps have specific characteristics?}
\label{res:charac}

\textbf{Motivation.}
After categorizing \covidRelated apps from their descriptions, in-depth analysis is needed to better understand how they work compared to standard apps. This section aims at comparing \covidRelated apps and standard apps from a technical point of view to bring insight into future research. To that end, we extract Android apps-related features (e.g., GUI components, permissions, libraries, etc.). The outcome of this research question will provide the reader with detailed information about \covidRelated apps. Indeed, it gives information on whether \covidRelated apps are more prone to track and/or display advertisements than standard apps.

\noindent
\textbf{Strategy.}
In prior work, \citet{tian2015characteristics} have shown that specific 
sets of apps can have similar characteristics (e.g., similar permissions, components, size, etc.).
In this section, we investigate to what extent \covidrelatedApps{} apps 
form one coherent group that is significantly 
different than other apps.

To that end, for each app, we counted the number of different Android components (i.e., 
\texttt{Activities}, \texttt{Broadcast Receivers}, \texttt{Services}, and \texttt{Content Providers}), computed the size of the dex file, extracted the permissions needed as well as the libraries used.

\textbf{Comparison Dataset:} 

For comparing the characteristics of \covidRelated apps with other apps characteristics, we randomly selected 100 apps over 10 different categories of apps from \gp{}. Those 1000 (10 x 100) apps are sampled from the same time span (i.e., they are coming from the same initial dataset) to ensure that time is not a factor in potential differences. 

\gp{} contains dozens of categories, therefore we decided to compare our set of \covidRelated apps against apps from the categories that intersect those of our \covidRelated apps. Table~\ref{table:categories} shows the categories of \covidRelated apps we were able to retrieve. Note that we were able to get the category of 87 among our set of \covidrelatedApps{} \covidRelated apps.

\begin{table}[ht]
    \centering
    \begin{tabular}{l|r|l|r}
    Category & \# of apps & Category & \# of apps \\ 
    \hline
    Communication & 2 & Entertainment & 1 \\
    Productivity & 1 & Medical & 23 \\
    Health \& Fitness & 48 & Tools & 3 \\
    Social & 3 & Lifestyle & 4 \\
    Shopping & 1 & Travel \& Local & 1 \\
    \end{tabular}
    \caption{Categories of \covidRelated apps and the number of apps in each category}
    \label{table:categories}
\end{table}

\begin{figure}[ht]
    \begin{subfigure}[b]{.49\linewidth}
        \centering
        \includegraphics[width=\linewidth]{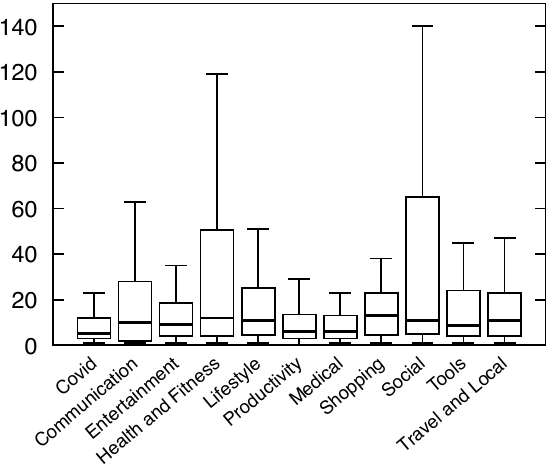}
        \caption{\# Activities}
    \end{subfigure}
    \begin{subfigure}[b]{.49\linewidth}
        \centering
        \includegraphics[width=\linewidth]{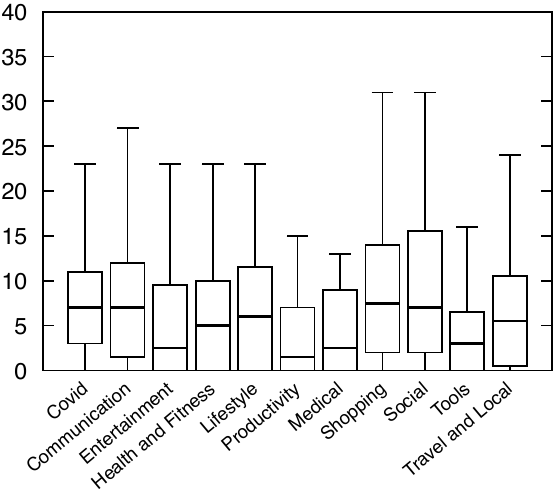}
        \caption{\# Services}
    \end{subfigure}
    \begin{subfigure}[b]{.49\linewidth}
        \centering
        \includegraphics[width=\linewidth]{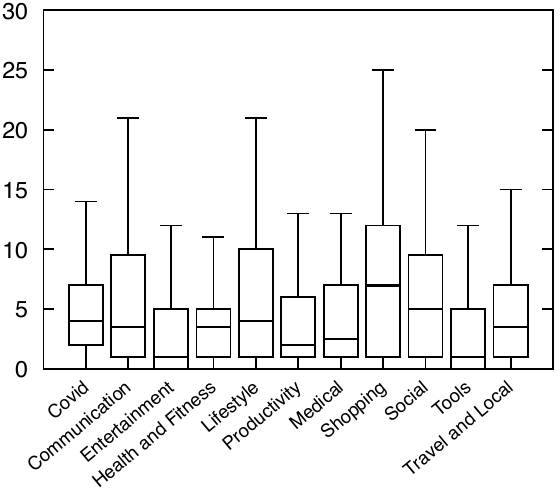}
        \caption{\# Receivers}
    \end{subfigure}
    \begin{subfigure}[b]{.49\linewidth}
        \centering
        \includegraphics[width=\linewidth]{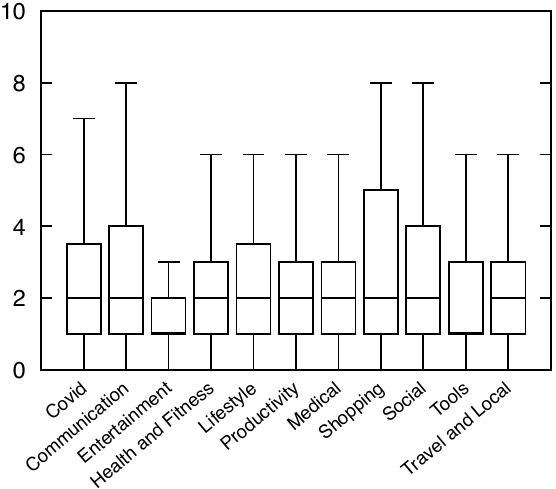}
        \caption{\# Providers}
    \end{subfigure}
    \caption{Number of components: Comparison between apps in categories and \covidRelated apps.}
    \label{fig:components}
\end{figure}

\textbf{Android Components:} 

Figure~\ref{fig:components} depicts differences between apps in different categories and our set of \covidRelated apps regarding the number of components included in the app. We notice that \covidRelated apps tend to use fewer Activities than the other apps. This difference is statistically confirmed to be significant by a  Mann-Whitney-Wilcoxon (MWW~\footnote{All MWW tests are done with significance level set at \num{0.05}}) test~\citep{mann1947,Wilcoxon1945}). Regarding Services (used for background tasks), we can see that, apart from the category ”Shopping”, \covidRelated apps tend to use more service components. Regarding broadcast receivers, however, the difference is less marked, although its statistical significance is confirmed by a MWW test. Finally, the median number of Content providers in \covidRelated apps is in most cases equal to the median number of Content providers of apps in different categories (i.e., 2 content providers). An MWW  test found no statistically significant difference (except for the \textit{Entertainment} category). 

Overall, the differences, which are mostly pronounced for Activities, suggest that \covidRelated apps are different from other apps (of the same category) in terms of GUI layout. With less Activities, we can conclude that \covidRelated may have less complex GUI than other apps. Services being slightly more used in \covidRelated apps, it hints that \covidRelated apps are more data-centric than other apps (in the same categories).

\textbf{Dex files size:}

\begin{figure}[ht]
    \centering
    \captionsetup{justification=centering}
    \includegraphics[width=0.6\linewidth]{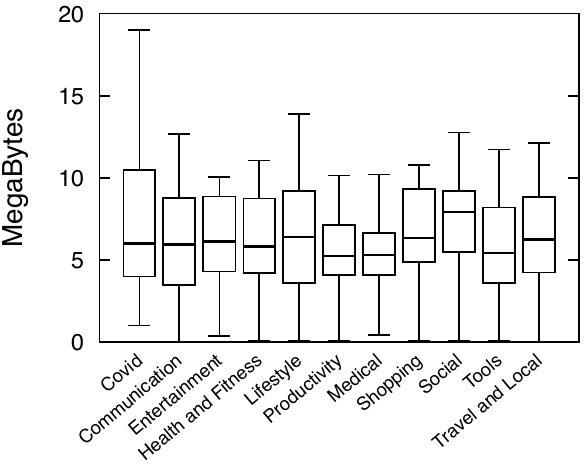}
    \caption{Size of applications: Comparison between apps in categories and \covidRelated apps}
    \label{fig:sizes}
\end{figure}

Figure~\ref{fig:sizes} shows the distributions of the dex sizes of \covidRelated apps and the apps from the ten different categories. It shows that the median of \covidRelated apps sizes is close to other apps in general. The MWW test confirms  no statistically significant difference between the distributions of app size. However, the maximum dex size value is higher than other apps, hinting at more variability in terms of app size amongst \covidRelated apps.

\textbf{Permissions:} 

In Table~\ref{table:permissions}, we compare the permissions used by \covidRelated apps and the permissions of other apps per app category. To that end, we extracted for all sets of apps the top ten most requested permissions. First, a notable difference is that \covidRelated apps tend to use the \texttt{wake\_lock} permission more than standard apps. This permission is used for preventing the screen of the device from being turned off, and/or to ensure an app remains active. Such a feature is often used for keeping the phone awake while locating the phone (e.g., for contact tracing). In the same way, \texttt{access\_fine\_location} and \texttt{access\_coarse\_location} tend to be used more by \covidRelated apps. This is in line with the use of the \texttt{wake\_lock} permission to facilitate user location tracking.

\begin{table}[ht]
    \centering
    \begin{adjustbox}{width=\textwidth}
    \begin{tabular}{llrlrlr}
        \hline
        \rowcolor{gray!70}
        \multicolumn{7}{c}{\textbf{\LARGE Permissions}} \\ \hline \hline
        \rowcolor{gray!40}
        & \multicolumn{1}{c}{\textbf{Communication}} & \textbf{\%} & \multicolumn{1}{c}{\textbf{Entertainment}} & \textbf{\%}  & \multicolumn{1}{c}{\textbf{Health \& Fitness}} & \textbf{\%} \\ \hline
        1 & INTERNET & 100\% & INTERNET & 98\% & INTERNET & 99\% \\ \hline
        2 & ACCESS\_NETWORK\_STATE & 97\% & ACCESS\_NETWORK\_STATE & 98\% & ACCESS\_NETWORK\_STATE & 90\% \\ \hline
        3 & WRITE\_EXTERNAL\_STORAGE & 90\% & WRITE\_EXTERNAL\_STORAGE & 74\% & WRITE\_EXTERNAL\_STORAGE & 77\% \\ \hline
        4 & READ\_EXTERNAL\_STORAGE & 77\% & ACCESS\_WIFI\_STATE & 69\% & WAKE\_LOCK & 67\% \\ \hline
        5 & WAKE\_LOCK & 76\% & READ\_EXTERNAL\_STORAGE & 57\% & ACCESS\_WIFI\_STATE & 67\% \\ \hline
        6 & c2dm.permission.RECEIVE & 71\% & WAKE\_LOCK & 51\% & READ\_EXTERNAL\_STORAGE & 57\% \\ \hline
        7 & ACCESS\_WIFI\_STATE & 62\% & RECEIVE\_BOOT\_COMPLETED & 36\% & c2dm.permission.RECEIVE & 53\% \\ \hline
        8 & VIBRATE & 61\% & ACCESS\_FINE\_LOCATION & 30\% & ACCESS\_COARSE\_LOCATION & 53\% \\ \hline
        9 & CAMERA & 61\% & c2dm.permission.RECEIVE & 29\% & ACCESS\_FINE\_LOCATION & 52\% \\ \hline
        10 & RECEIVE\_BOOT\_COMPLETED & 49\% & ACCESS\_COARSE\_LOCATION & 29\% & RECEIVE\_BOOT\_COMPLETED & 41\% \\ \hline
    \end{tabular}
    \end{adjustbox}
    
    \begin{adjustbox}{width=\textwidth}
    \begin{tabular}{llrlrlr}
        \hline    
        \rowcolor{gray!40}
        & \multicolumn{1}{c}{\textbf{Productivity}} & \textbf{\%} & \multicolumn{1}{c}{\textbf{Medical}} & \textbf{\%}  & \multicolumn{1}{c}{\textbf{Shopping}} & \textbf{\%} \\ \hline
        1 & INTERNET & 97\% & INTERNET & 98\% & INTERNET & 100\% \\ \hline
        2 & ACCESS\_NETWORK\_STATE & 93\% & ACCESS\_NETWORK\_STATE & 92\% & ACCESS\_NETWORK\_STATE & 97\% \\ \hline
        3 & WRITE\_EXTERNAL\_STORAGE & 90\% & WRITE\_EXTERNAL\_STORAGE & 73\% & WRITE\_EXTERNAL\_STORAGE & 84\% \\ \hline
        4 & ACCESS\_WIFI\_STATE & 61\% & WAKE\_LOCK & 55\% & WAKE\_LOCK & 77\% \\ \hline
        5 & READ\_EXTERNAL\_STORAGE & 54\% & READ\_EXTERNAL\_STORAGE & 51\% & c2dm.permission.RECEIVE & 75\% \\ \hline
        6 & WAKE\_LOCK & 53\% & ACCESS\_WIFI\_STATE & 49\% & READ\_EXTERNAL\_STORAGE & 68\% \\ \hline
        7 & CAMERA & 50\% & c2dm.permission.RECEIVE & 48\% & ACCESS\_WIFI\_STATE & 64\% \\ \hline
        8 & c2dm.permission.RECEIVE & 44\% & CAMERA & 44\% & VIBRATE & 55\% \\ \hline
        9 & ACCESS\_FINE\_LOCATION & 43\% & VIBRATE & 41\% & RECEIVE\_BOOT\_COMPLETED & 54\% \\ \hline
        10 & ACCESS\_COARSE\_LOCATION & 38\% & ACCESS\_FINE\_LOCATION & 39\% & CAMERA & 54\% \\ \hline
    \end{tabular}
    \end{adjustbox}
    
    \begin{adjustbox}{width=\textwidth}
    \begin{tabular}{llrlrlr}
        \hline
        \rowcolor{gray!40}
        & \multicolumn{1}{c}{\textbf{Lifestyle}} & \textbf{\%} & \multicolumn{1}{c}{\textbf{Tools}} & \textbf{\%}  & \multicolumn{1}{c}{\textbf{Travel \& Local}} & \textbf{\%} \\ \hline
        1 & INTERNET & \num{99}\% & INTERNET & 97\% & INTERNET & 100\% \\ \hline
        2 & ACCESS\_NETWORK\_STATE & \num{97}\% & ACCESS\_NETWORK\_STATE & 91\% & ACCESS\_NETWORK\_STATE & 98\% \\ \hline
        3 & WRITE\_EXTERNAL\_STORAGE & \num{78}\% & WRITE\_EXTERNAL\_STORAGE & 76\% & WRITE\_EXTERNAL\_STORAGE & 84\% \\ \hline
        4 & ACCESS\_WIFI\_STATE & \num{69}\% & ACCESS\_WIFI\_STATE & 60\% & WAKE\_LOCK & 76\% \\ \hline
        5 & WAKE\_LOCK & \num{67}\% & READ\_EXTERNAL\_STORAGE & 55\% & ACCESS\_FINE\_LOCATION & 70\% \\ \hline
        6 & c2dm.permission.RECEIVE & \num{55}\% & WAKE\_LOCK & 48\% & ACCESS\_WIFI\_STATE & 68\% \\ \hline
        7 & READ\_EXTERNAL\_STORAGE & \num{55}\% & ACCESS\_FINE\_LOCATION & 42\% & ACCESS\_COARSE\_LOCATION & 64\% \\ \hline
        8 & ACCESS\_FINE\_LOCATION & \num{46}\% & ACCESS\_COARSE\_LOCATION & 40\% & c2dm.permission.RECEIVE & 63\% \\ \hline
        9 & VIBRATE & \num{44}\% & VIBRATE & 34\% & READ\_EXTERNAL\_STORAGE & 60\% \\ \hline
        10 & ACCESS\_COARSE\_LOCATION & \num{42}\% & c2dm.permission.RECEIVE & 31\% & VIBRATE & 41\% \\ \hline
    \end{tabular}
    \end{adjustbox}
    \begin{adjustbox}{width=.76\textwidth}
    \begin{tabular}{llrlr}
        \hline
        \rowcolor{gray!40}
        & \multicolumn{1}{c}{\textbf{COVID}} & \textbf{\%} & \multicolumn{1}{c}{\textbf{Social}} & \textbf{\%} \\ \hline
        1 & INTERNET & \num{98.86}\% & INTERNET & 99\% \\ \hline
        2 & ACCESS\_NETWORK\_STATE & \num{93.18}\% & ACCESS\_NETWORK\_STATE & 96\% \\ \hline
        3 & WAKE\_LOCK & \num{78.40}\% & WRITE\_EXTERNAL\_STORAGE & 87\% \\ \hline
        4 & c2dm.permission.RECEIVE & \num{73.86}\% & READ\_EXTERNAL\_STORAGE & 80\% \\ \hline
        5 & ACCESS\_FINE\_LOCATION & \num{65.90}\% & WAKE\_LOCK & 76\% \\ \hline
        6 & BIND\_GET\_INSTALL\_REFERRER\_SERVICE & \num{59.10}\% & ACCESS\_WIFI\_STATE & 73\% \\ \hline
        7 & ACCESS\_COARSE\_LOCATION & \num{54.64}\% & c2dm.permission.RECEIVE & 69\% \\ \hline
        8 & RECEIVE\_BOOT\_COMPLETED & \num{50.00}\% & ACCESS\_COARSE\_LOCATION & 63\% \\ \hline
        9 & FOREGROUND\_SERVICE & \num{50.00}\% &  ACCESS\_FINE\_LOCATION & 61\% \\ \hline
        10 & WRITE\_EXTERNAL\_STORAGE & \num{38.64}\% & CAMERA & 50\% \\ \hline
    \end{tabular}
    \end{adjustbox}
    \caption{Top ten most requested permissions in \covidRelated apps and other apps per category. Percentage indicates the ratio of apps using
    the permission.}
    \label{table:permissions}
\end{table}

\begin{figure}[ht]
    \centering
    \captionsetup{justification=centering}
    \includegraphics[width=0.6\linewidth]{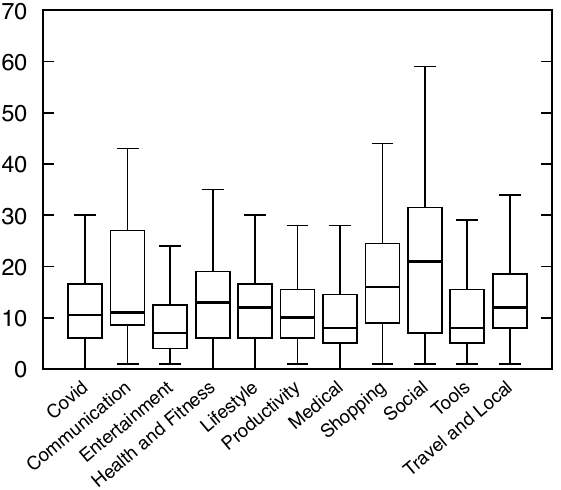}
    \caption{Number of permissions per app.}
    \label{fig:permissions}
\end{figure}

Figure~\ref{fig:permissions} shows the distribution of the number of permissions requested by \covidRelated apps and other apps per category. The MWW tests revealed no statistically significant difference between the number of permissions used by \covidRelated apps and by other apps

\textbf{Libraries:} To compare the patterns of libraries inclusion, 
we measure the use of libraries by relying on a collection of well-known libraries.
More specifically, we re-use two lists of libraries established in prior works~\citep{LI2019157,LiSaner2016}: 
a list of 1\,114 common libraries and a list of 

240 advertisement libraries.
Therefore, for \covidRelated apps and our dataset of apps by category, we computed the number of apps using at least one common library and one advertisement library.

Table~\ref{table:libraries} presents our results.
First, we notice that almost all the apps (\covidRelated and other) use common
libraries, which is not surprising since Android software development---just like 
non-mobile software---heavily relies on reusable libraries and frameworks.

However, the difference is significant regarding the advertisement libraries. 
Indeed, while advertisement libraries are used by more than 80\% of 
other apps, they only appear in less than 20\% of \covidRelated apps.
Furthermore, only 3 out of 240 advertisement libraries are used in 
\covidRelated apps, namely: (1) \emph{com.facebook}, (2) \emph{com.startapp.android}
and (3) \emph{com.flurry}.
This strongly suggests that the primary goal of \covidRelated apps is not to 
obtain a financial gain from advertisement, in opposition to the vast majority of standard apps.

\begin{table}[ht]
    \centering
    \begin{adjustbox}{width=\textwidth}
    \begin{tabular}{l|r|r|r|r|r|r|r|r|r}
        Type of library & COVID & C & E & H\&F & L & P & M & SP & S \\ 
        \hline
        Common Library & 100\% & 96\% & 98\% & 93\% & 95\% & 97\% & 97\% & 98\% & 93\% \\
        \hline
        Ad Library & 19.6\% & 93\% & 98\% & 87\% & 92\% & 83\% & 86\% & 93\% & 86\% \\
    \end{tabular}
    \end{adjustbox}
    \caption{Number of \covidRelated/other apps using libraries. (C: Communication, E: Entertainment, H\&F: Health \& Fitness, L: Lifestyle, P: Productivity, M: Medical, SP: Shopping, S: Social)}
    \label{table:libraries}
\end{table}

\highlight{
    \textbf{RQ 2 Answer:} \covidRelated apps are composed of less
    GUI-related components than other apps.
    Their size and the number of permissions they use are not different from other apps.
    They tend to use more tracking-related permissions.
    With respect to libraries, \covidRelated apps use 
    common libraries just like any other apps, but significantly fewer advertisement libraries than other apps.
    We characterize the differences/similarities between \covidRelated apps and standard apps which brings insight for future research.
    End-users can take note that \covidRelated apps could track them and that they are not likely to use advertisement libraries.
}

\subsection{Are \covidRelated Android apps more complex than standard apps?}
\label{complexity}

\textbf{Motivation.}
In Section~\ref{covid_used_for}, we have seen that \covidRelated apps cover a large variety of categories and target various objectives (e.g., informing users, collecting data from users, etc.). The code complexity that is necessary to achieve these objectives may thus vary substantially. To investigate this aspect, we compute several standard metrics used in the state of the art literature, and further assess the potential differences between \covidRelated apps and other apps. Insights from this research question can improve developer’s knowledge and serve as the basis for future empirical research on code quality.

\noindent
\textbf{Strategy.}
App complexity is an elusive concept. Yet, in the literature, there are various studies that 
propose metrics to measure some form of complexity and attempt to show its correlation 
with app quality and maintainability~\citep{jovst2013using,gao2019evolution}. 
We undertake to investigate our research question based on these common metrics from the literature~\citep{chidamber1994metrics}.
We provide in Appendix~\ref{appendix:complexity} the descriptions of the complexity metrics 
we use.

In this study, the data extracted for computing the complexity metrics are computed at the smali code level. The apps are loaded with Androguard~\citep{androguard}, a static analysis tool for Android apps.

The different metrics attempt to capture the \emph{Lack of Cohesion in Methods} (LCOM), the \emph{Weighted number of Methods per Class} (WMC), the number of methods invoked per class, i.e., the \emph{Response For a Class} (RFC), the \emph{Coupling Between Object classes} (CBO) and the \emph{Number Of Children per class} (NOC).

\begin{figure}[H]
    \begin{subfigure}[b]{.49\linewidth}
        \centering
        \includegraphics[width=\linewidth]{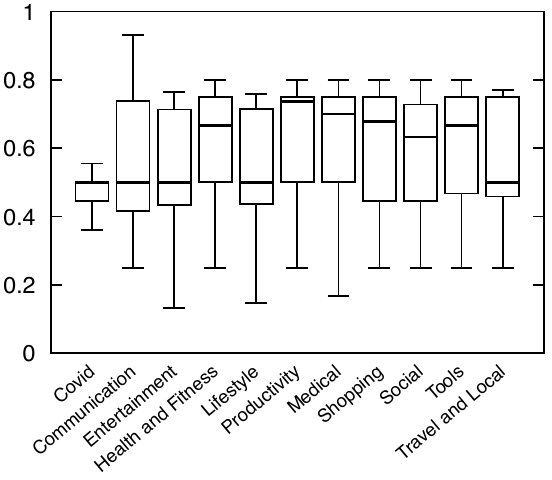}
        \caption{LCOM}
    \end{subfigure}
    \begin{subfigure}[b]{.49\linewidth}
        \centering
        \includegraphics[width=\linewidth]{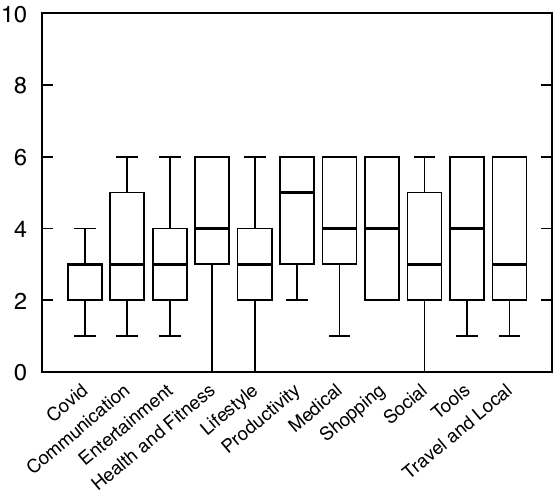}
        \caption{WMC}
    \end{subfigure}
    \begin{subfigure}[b]{.49\linewidth}
        \centering
        \includegraphics[width=\linewidth]{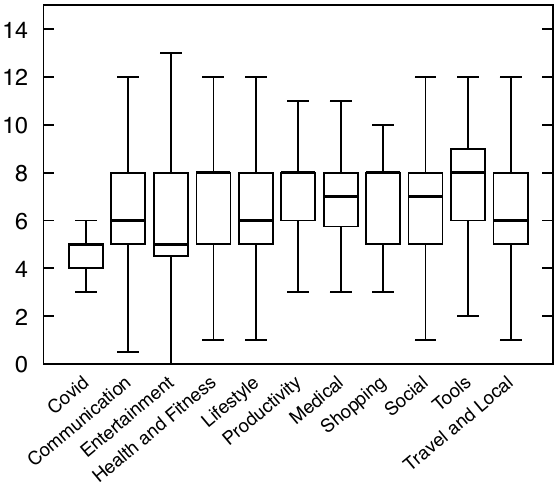}
        \caption{RFC}
    \end{subfigure}
    \begin{subfigure}[b]{.49\linewidth}
        \centering
        \includegraphics[width=\linewidth]{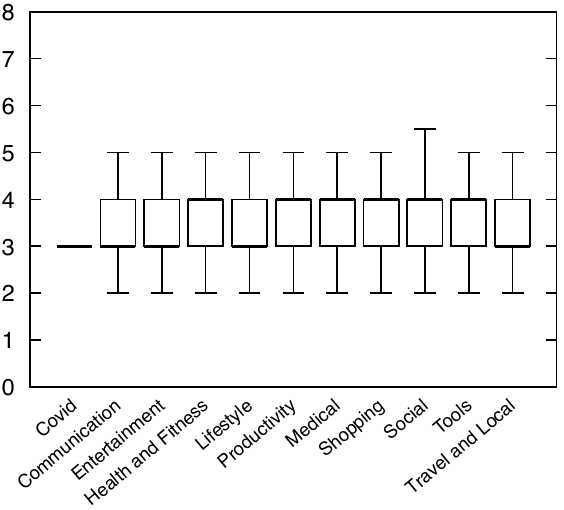}
        \caption{CBO}
    \end{subfigure}
    \begin{subfigure}[b]{.49\linewidth}
        \centering
        \includegraphics[width=\linewidth]{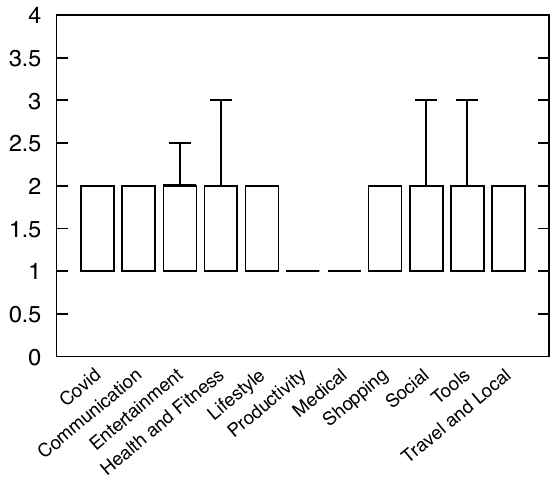}
        \caption{NOC}
    \end{subfigure}
    \caption{Complexity of \covidRelated apps compared to standard apps.}
    \label{fig:complexity}
\end{figure}

Figure~\ref{fig:complexity} presents the distributions of metric values.

NOC appears to present similar distribution across standard and \covidRelated apps, confirmed by MWW test.
However, MWW test revealed significant differences between the distributions of \covidRelated apps and standard apps for the other metrics~\footnote{WMC ($\bar{p} = \num{3.48e-05}$), RFC ($\bar{p} = \num{5.93e-06}$), CBO ($\bar{p} = \num{1.1e-04}$) and LCOM ($\bar{p} = \num{1.5e-02}$)}.

Furthermore, Figure~\ref{fig:complexity} distinctly shows that \covidRelated apps complexity metric medians' are below standard apps medians which hints at a lower complexity.

We note that obfuscation is a factor that can have an impact on Android apps studies, especially with app complexity computation based on smali code.
Our set of \covidRelated apps contains 2 apps (\num{2.17}\%) that contain obfuscated code. For measuring if an app uses obfuscated code, we rely on APKiD~\footnote{https://github.com/rednaga/APKiD}.
The obfuscation rate of apps in each category is depicted in Table~\ref{table:obfuscation_rate}.

\begin{table}[ht]
    \centering
    \begin{tabular}{l|r}
    Category & Obfuscated apps rate \\ 
    \hline
    Covid & 2.17\% \\
    Communication & 18\% \\
    Entertainment & 20\% \\
    Health \& Fitness & 15\% \\
    Lifestyle & 16\% \\
    Medical & 3\% \\
    Productivity & 5\% \\
    Shopping & 27\% \\
    Social & 36\% \\
    Tools & 34\% \\
    Travel \& Local & 50\% \\
    \end{tabular}
    \caption{Rate of apps obfuscated by category}
    \label{table:obfuscation_rate}
\end{table}

At first sight, we can see in Table~\ref{table:obfuscation_rate} that for some categories, there is a high number of apps that contain obfuscated code, which suggests that the metrics computed can be biased by the obfuscation rates.

We therefore conducted the same comparisons, but based on random sets of non-obfuscated apps. The conclusions remain the same.

Overall, these results establish that \covidRelated apps are, to some extent, less complex than standard apps.  
According to~\citep{jovst2013using}, this result suggests that \covidRelated apps may be more maintainable and of better quality.
Additionally, we note that a lower complexity could also indicate that \covidRelated apps have on average less 
functionalities and/or are focused on more specific goals, as was already hinted above in 
the permission usages comparison.

\highlight{
    \textbf{RQ 3 Answer:} Our empirical study shows that \covidRelated apps tend to be less complex than standard apps. We also show that obfuscation is not a factor impacting the complexity of \covidRelated apps in our study dataset.
}
    
\subsection{To what extent were \covidRelated apps removed from
        the official \gp{} and why?}
\label{remove_playstore}

\textbf{Motivation.}
Developers have to comply with strict Google policies~\citep{googlepolicies} before submitting an app to \gp{}. The unprecedented crisis of the \covid led Google to release new policies regarding \covidRelated apps that would be candidates for \gp{}~\citep{googleupdate} where Google performs supplementary checks (e.g., reduce misinformation by favoring official sources). With this research question, we aim to check to what extent Google actually applied strict policies in \gp{}. The outcome of this research question will open new research avenues for app policy modeling, and may shed light into Google vetting processes for developers.

\noindent
\textbf{Strategy.}
We have seen in Section~\ref{covid_used_for} that during our analyses, some \covidRelated apps disappeared from the official \gp{} in a matter of days.

Therefore, for each app that was initially identified at the beginning of our study, we queried the
\gp{} market, at the time of writing, to check if the app is still available. 
Around 15\% of \covidRelated apps (i.e., \covidAppsNotAvailableDesc{} apps) have been removed from \gp{}.

In comparison, among \num{1675} standard apps taken randomly from our initial dataset
(see Section~\ref{background_dataset}), we found that 277 (i.e. \num{16.54}\%)
apps were removed from the \gp{} market.

The removal rates of both app datasets are close.
Actually, we expected a much higher removal rate for \covidRelated apps.
This relatively low ratio of removal for \covidRelated apps could be explained in several ways:
\begin{itemize}
    \item Google either enforces its policy very quickly or pre-screens (i.e., before it is accepted on the market) each app that is potentially relevant to \covid;
       In that case, apps would either never make it to the market, or would be removed too quickly 
       for \az{} crawlers to catch them;
    \item App developers either rapidly adapted to Google's policy and/or very few developers 
    proposed apps that conflict with Google's policy.
\end{itemize}

\highlight{
    \textbf{RQ 4 Answer:}
    The data that we were able to collect was not sufficient to reliably answer the research question. Although we could conclude that \covidRelated apps are removed from \gp{} at a normal rate, we could not conclude on which policies were applied to justify removal from the market. Furthermore, the lack of data about the stream of apps that are submitted to \gp{} validation process (i.e., even prior to publication on the market) prevents a reliable investigation on the application of policies. Nevertheless, our findings should remind developers of \covidRelated apps that their apps will be subject to policy checks. Our inconclusive findings about the applied policies further suggest that market policy inference constitutes a research challenge for the Android community.
}

\subsection{Who are \covidRelated apps' developers?}
\label{developers}

\textbf{Motivation.}
In Section~\ref{res:dates}, we have seen that the number of \covidRelated apps increased drastically from March 2020. The important information behind this is that many entities quickly responded to the pandemic to provide users with specific Android apps with different purposes (e.g., information, contact tracing, health guide, etc.). However, the nature of the entities was not readily available. In this section, we consider further investigating their type by mining description data and following various links. Typically, we focus on retrieving the origin of those apps to overview what country responded according to the pandemic to provide services to end-users. This information would help to overview which countries quickly reacted to the pandemic by providing end-users with mobile apps services. The outcome of this research question will give the general public a glimpse of the distribution of apps by country. Besides, by exposing the type of developers and the origin of \covidRelated apps, we encourage future research into performing additional studies such as code reuse in different apps/countries, plagiarism between apps, as well as correlation between app releases and the number of \covid cases.

\noindent
\textbf{Strategy.}
On \gp{}, in each web page of an app\footnote{https://play.google.com/store/apps/details?id=PACKAGENAME, where PACKAGENAME is the package name of the app.},  
there is a field $developer$ that provides the name  of the person or entity (e.g. a software company, a governmental institution, an ONG, etc.) who has released the app. 
After collecting this information, we detail in Table~\ref{first_part_table} (column \emph{Developer Type}) the status (or the type) of the entity having released an app. 
Table~\ref{table:numberOfEntities} presents the number of released \covidRelated apps for each type of entities.

\begin{table}[ht]
    \centering
    \begin{tabular}{l|r|r}
    	Type of the entity & \# of apps & \%\\ 
    	\hline
      Governmental Institution & 52 & 66.67\%\\ 
        Company  & 17 & 21.79\%\\
        Association & 3 & 3.85\%\\
Independent & 2 & 2.56\%\\
Researchers & 2 & 2.56\%\\
NGO & 1& 1.28\%\\
Hospital & 1& 1.28\%\\
\hline
Total & 78 & 100\%
    \end{tabular}
    \caption{Number of \covidRelated apps per entity type }
    \label{table:numberOfEntities}
\end{table}

We can see that most of the app providers are
governmental institutions. 
We indeed find \covidRelated apps that are officially promoted by national governments (e.g. Government of Brazil\footnote{br.gov.datasus.guardioes} or Government of France\footnote{fr.gouv.android.stopcovid}).
We also see apps released by more local governmental bodies (at the state or regional level). 
We have for instance apps from specific states of the USA (e.g., State of Rhode Island\footnote{com.ri.crushcovid}), 
or from specific "Switzerland Canton" (e.g. Gesundheitsdepartement des Kantons Basel-Stadt\footnote{ch.covid19bs.app.PMSMobile}).

About 20\% of the \covidRelated apps (17 apps) are provided by companies. 
In order to understand why these apps have not been removed by Google, we further check the description of these apps and the descriptions of the companies. 
We found that:
\begin{itemize}
    \item Even if the developer is identified as a company, two apps have been developed on behalf of official bodies (\emph{Care19}\footnote{com.proudcrowd.care} is the official COVID-19 app for the states of South Dakota and North Dakota, \emph{COVID AP-HM}\footnote{com.ambulis.aphm.covid} is an app developed for a hospital); 
    
    \item Seven apps are either endorsed by a ministry\footnote{covid19care.virus.coronavirus.corona.sick.marcom.health.pakistan},
or working in close collaboration with medical/health actors\footnote{com.docandu.checker, it.adl.aslroma3.covid19.app, it.adilife.covid19.app, com.maithu.transplantbuddy.covid19, de.bssd.covid19}, 
or working in collaboration with renowned universities\footnote{com.joinzoe.covid\_zoe}.

    \item Two apps are actually online shopping apps\footnote{com.coronavirus.facemask  and com.krrsoftwaresolutions11.Facemasks}.

    \item One app is not on the market anymore\footnote{com.dawsoftware.contacttracker}.

    \item Finally, five apps related to social distancing\footnote{world.coalition.app and com.bloomreality.sodi }, or health\footnote{de.kreativzirkel.coronika and it.softmining.projects.covid19.savelifestyle}, or Covid-related news\footnote{com.osapps.covid19}, 
    have been released by companies without any explicit link to official organizations. 
    We remind that the official Google \covid  policy~\citep{googleupdate} is that \covidRelated apps with no explicit links with governmental bodies or health organizations cannot provide "health claims". 
We further check these 5 apps, and we confirm that they comply with the Google \covid policy.

\end{itemize}

For the remaining nine \covidRelated apps, we noticed that 3 apps have been provided by associations.
More specifically, the \emph{DiagnoseMe}\footnote{bf.diagnoseme.fasocivic} app has been released by the Faso Civic association from Burkina Faso, 
the \emph{Self Shield App}\footnote{org.sshield.selfshield} by the Commonwealth Medical Association (through the Commonwealth Centre for Digital Health organization) and the \emph{COVID Safe Paths}\footnote{org.pathcheck.covidsafepaths} app by a non-profit organization related to MIT.
We also noticed that two apps have been developed by independent developers, and two other apps have been provided by researchers. One by a group of researchers from German Universities\footnote{com.coronacheck.haugxhaug.testyourcorona}, one by researchers from the Aga Khan University in Pakistan\footnote{com.edu.aku.akuhccheck}.  
Finally, one app has been provided by an NGO (i.e., the Austria Red Cross), and one by a hospital (actually a group of hospitals in Paris, France).

We note that among all the \covidRelated apps, 
71\% of them have been released by entities having multiple Android apps on \gp{}.

\begin{figure}[ht]
    \centering
    \includegraphics[width=\textwidth]{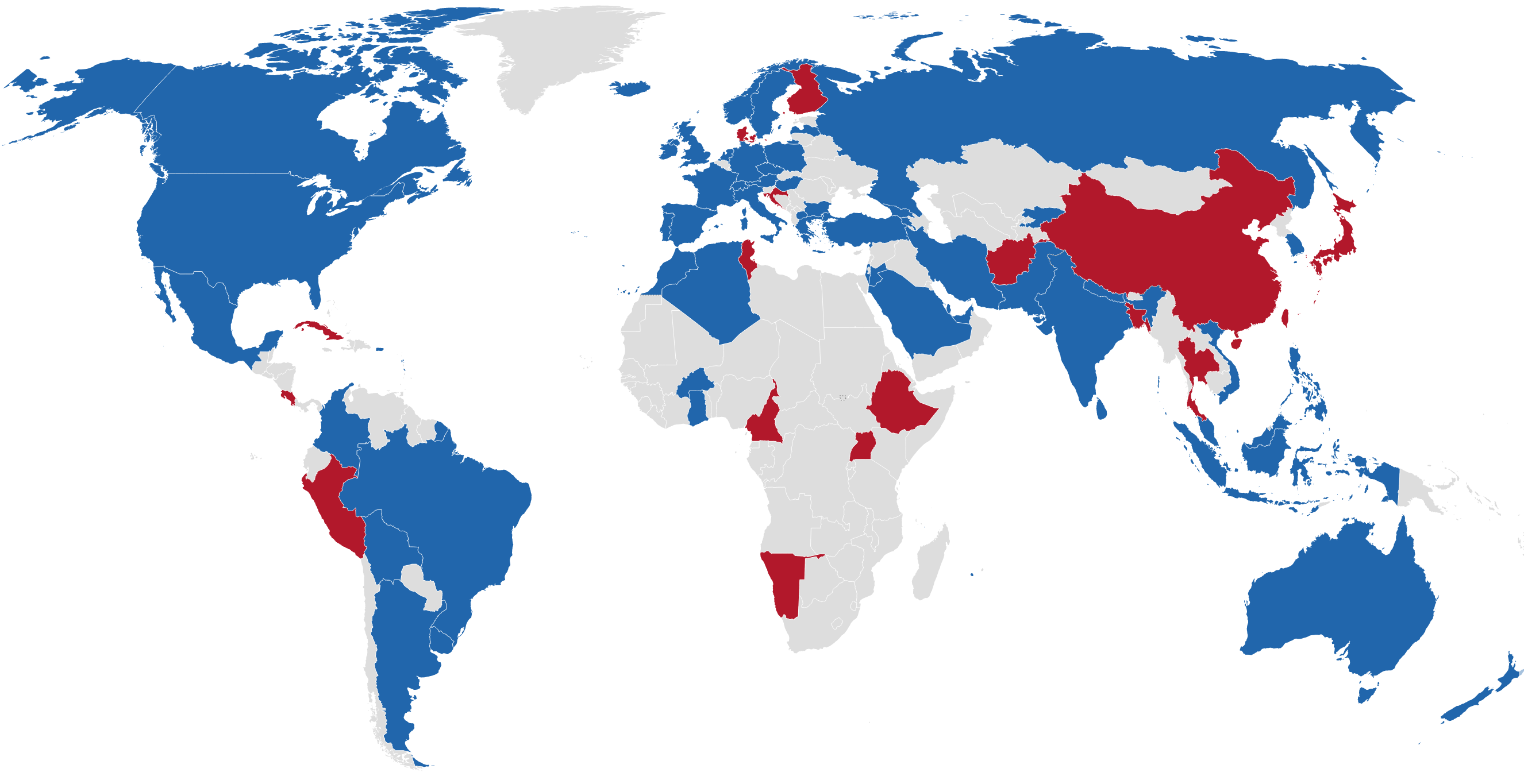}
    \caption{Countries of origin of \covidRelated apps (Blue: Apps available, Red: Apps not available)}
    \label{fig:map}
\end{figure}

Finally, we represent in the map of Figure~\ref{fig:map} the geographical distribution of the apps over the world. 
We can see that \covidRelated apps are provided world-wide (maybe less present in Africa).
The countries in blue are the ones listed in Table~\ref{first_part_table}.
Note that we also identified \countriesNotAvailable{} other apps from \countriesNotAvailable{} 
countries that we were unable to obtain; These countries are represented in red.

\highlight{
    \textbf{RQ 5 Answer:}
    Most \covidRelated apps that are seen on \gp{} are developed with, and for official government entities or public health organizations. We note that apps developed by non-official entities are not making potentially misleading health claims, as per Google’s policy. \covidRelated apps have been released in countries from all continents. The majority (71\%) of \covidRelated app developers have at least one other app available on \gp{}. We provide useful insights for the research field of \covidRelated apps by exposing the type of developers who produce \covidRelated apps. Developers located in countries that have not yet developed a \covidRelated app could benefit from this information by having a clear overview of what other countries are doing, hence informing their choice of potentially developing their own \covidRelated app to fight against the virus. The broad-public will be able to easily see whether an app has been developed for their country, and the entity that supervised its development.
}

\subsection{Do \covidRelated apps have security issues?}
\label{malicious}

\textbf{Motivation.}
Security and privacy are critical concerns regarding mobile apps. In this section, we assess several aspects of \covidRelated apps security. The outcome of this research question will provide the general public with a summary of some potential security problems found in \covidRelated apps, which may help them adjust their level of trust in such apps. Similarly, we highlight potential issues that developers should consider in terms of security of \covidRelated apps. Researchers may also use this information to adopt further investigation topics related to \covidRelated apps security, e.g., on API usage patterns, evolution of security and privacy in the lineage of apps, etc.

\noindent
\textbf{Strategy.}
In contrast to a recent work~\citep{he2020virus}, which focused on dissecting \covidRelated malware, our aim in this work is not to perform an extensive security analysis of these apps. Nevertheless, we propose to leverage four practical security and privacy scanners on our set of \covidrelatedApps{} \covidRelated apps in order to systematically evaluate four S\&P aspects: 
(1) the presence of privacy leaks; 
(2) the number of apps flagged by \vt{};
(3) the misuse of crypto-APIs;
(4) the matching between descriptions and behavior.

\textbf{[Privacy leaks]} 
As we have seen in~\ref{covid_used_for}, most of the \covidRelated apps are made for collecting personal and sensitive data, e.g., health data and/or the location of users.
Therefore, the security and privacy aspects of these apps are crucial, and many people started to share
concerns related to this topic~\citep{covid19privacyproblem,covid19privacyproblem1,covid19privacyproblem2}. 
In order to assess the privacy of \covidRelated apps, we applied the state-of-the-art data leak detector FlowDroid-IccTA~\citep{arzt2014flowdroid,li2015iccta}. 
Through static analysis, this tool is able to detect sensitive data leaks intra-component (e.g., inside an Activity) or inter-component (e.g., across Activities).
Note that we used the default sources and sinks provided with the tool.

FlowDroid-IccTA was able to detect 24 intra-component data leaks in 2 different apps and found no inter-component leak for the list of the 24 leaks.
The app \emph{SODI}\footnote{com.bloomreality.sodi} contained only 1 potential leak, whereas the app \emph{Coronavirus - SUS}\footnote{br.gov.datasus.guardioes} contained 23 potential leaks.
Given that static analysis tools are subject to false-positives, we undertake to manually analyze every detected leak.

We compiled the list of the 24 leaks in Table~\ref{table:leaks}. In the second column we expose the source of the potential leak, i.e., the sensitive information which is the first chain link. The third column lists the sinks associated with the sources, i.e., the method that is responsible for leaking the sensitive information.

\begin{table}[ht]
    \centering
    \begin{adjustbox}{width=\linewidth}
    \begin{tabular}{|r|l|l|}
    \hline
    \rowcolor{gray!70}
    & Sources & Sinks \\ \hline
    \rowcolor{gray!40}
    & \multicolumn{2}{c|}{com.bloomreality.sodi} \\ \hline
    1 & android.content.Context & android.content.Intent.registerReceiver \\ \hline
    \rowcolor{gray!40}
    & \multicolumn{2}{c|}{br.gov.datasus.guardioes} \\ \hline
    1 & android.content.Intent.getIntent & android.util.Log.w \\ \hline
    2 & android.os.Bundle & android.util.Log.w \\ \hline
    3 & android.content.Context & android.util.Log.d \\ \hline
    4 & android.content.Context & android.util.Log.d \\ \hline
    5 & java.net.URL.openConnection & android.util.Log.e \\ \hline
    6 & java.net.URL.openConnection & android.util.Log.e \\ \hline
    7 & android.location.Location.getLongitude & android.util.Log.d \\ \hline
    8 & org.apache.cordova.CordovaActivity.getIntent & android.util.Log.d \\ \hline
    9 & android.location.Location.getLongitude & android.util.Log.d \\ \hline
    10 & java.net.URL.openConnection & android.util.Log.d \\ \hline
    11 & java.net.URL.openConnection & android.util.Log.d \\ \hline
    12 & android.location.Location.getLatitude & android.util.Log.d \\ \hline
    13 & android.os.Bundle & android.util.Log.d \\ \hline
    14 & java.net.URL.openConnection & android.util.Log.d \\ \hline
    15 & android.location.Location.getLatitude & android.util.Log.d \\ \hline
    16 & android.content.Intent & android.util.Log.d \\ \hline
    17 & android.content.Intent & android.util.Log.d \\ \hline
    18 & com.google.firebase.messaging.RemoteMessage & android.util.Log.d \\ \hline
    19 & android.content.Intent & android.util.Log.d \\ \hline
    20 & android.content.Intent & android.util.Log.d \\ \hline
    21 & org.apache.cordova.CordovaActivity.getIntent & android.util.Log.d \\ \hline
    22 & android.os.Bundle & android.util.Log.d \\ \hline
    23 & java.net.URL.openConnection & android.util.Log.d \\ \hline
    \end{tabular}
    \end{adjustbox}
    \caption{List of the leaks detected by Flowdroid-IccTA. Note that there can be multiple leaks for each couple of source/sink.}
    \label{table:leaks}
\end{table}

\emph{SODI} an app promoting social-distancing. The app is not originating 
from a government.
Our manual analysis concluded, however, that the reported leak is a false-positive alarm and does not constitute a real data leak.

Regarding \emph{Coronavirus - SUS}, which is an official app of the government of 
Brazil, FlowDroid-IccTA flagged 24 potential sensitive leaks (i.e., there is a path between a 
source (that can access a sensitive data) to a sink (e.g. sendTextMessage)).
We notice that four of these leaks allow the app to get the longitude and/or latitude (the sources) of the app to log it internally (the sink).
However, this does not necessarily constitute a malicious behavior.

\textbf{[AntiVirus detection]}
For each of the \covidRelated apps, we have collected the detection reports from 
over 60 AntiVirus products, thanks to the VirusTotal API\footnote{\url{https://www.virustotal.com}}.
None of \covidRelated apps is flagged by any of the 60 anti-virus software available in VirusTotal at the time of writing.

\textbf{[Crypto-API misuses]}  
Finally, we leverage the state-of-the-art static-analyzer CogniCrypt~\citep{kruger2017cognicrypt} through its headless implementation CryptoAnalysis~\citep{cryptoanalysis} for detecting cryptographic API misuses in Java programs.
Such misuses could indeed indicate security issues. 
We found that 81 apps among our set of \covidrelatedApps{} \covidRelated apps use 
JCA\footnote{Java Cryptography Architecture, the most popular library of java Crypto-APIs} APIs.
However, CogniCrypt did not report any cryptographic misuse. 

In contrast, \citet{8816738} have shown that in a dataset of more than \num{598000} apks, 96\% of apks using JCA exhibit dangerous misuses of cryptographic APIs. 
With 0\%, \covidRelated apps seem to be totally exempt from such misuses. 

\textbf{[Description/Behavior matching]}
\covidRelated apps may propose functionalities that are sensitive due to the security and privacy concerns that they can raise (e.g., the  Indian app Aarogya Setu~\citep{Aarogya} was found to share users’ private information to third parties).
The apps we study have been released on \gp{}, therefore users can  only rely on the description provided by developers. 

However, it has been shown by~\citet{appbehavior} that apps’ behavior does not always match the apps’ description. 
For this reason, we replicated the CHABADA approach\citep{appbehavior} to check to what extent \covidRelated apps’ descriptions match their behavior (approximated by API usages). 
CHABADA unfolds as follows:

\begin{enumerate}
    \item Preprocessing descriptions with NLP techniques: tokenization, stop word removal, stemming
    \item Extracting topics with Latent Dirichlet Allocation~\citep{10.5555/944919.944937}
    \item Clustering apps based on topics with K-means~\citep{macqueen1967}
    \item Identifying, in each cluster, the apps that have outlier API usages. This outlier identification is performed via One-Class Support Vector Machine learning~\citep{doi:10.1162/089976601750264965}
\end{enumerate}

In section~\ref{covid_used_for}, we have seen that we were able to retrieve the descriptions of \covidrelatedAppsWithDesc{} apps. 
We applied our implementation of CHABADA to these \covidrelatedAppsWithDesc{} apps.
The clusters generated by CHABADA can be seen in Table~\ref{table:clusters}. 
Five clusters have been generated. We have named these clusters by considering the three most used words per cluster. 
We can see that the first cluster (i.e., Spread tracking) contains 30 apps, whereas other clusters are smaller and are all roughly the same size (i.e., between 10 and 14 apps per cluster). 
Do note that the clusters, which are independently built using the CHABADA approach, can each be associated to a category of our taxonomy (defined in Section~\ref{covid_used_for}).

After clustering the apps based on their description, CHABADA searches for outliers in each cluster based on the APIs usage. 
Table~\ref{table:clusters} shows the number of outliers detected per cluster. 
The three outliers in ”Spread tracking” have been detected because, contrary to other apps in the cluster, they use Android vibration API and Android MediaPlayer API. Similarly, in the cluster ”Sharing health information”, the two outliers use Bluetooth APIs, which is not the case for other apps in the cluster. 
Regarding the cluster ”Sharing general information”, the outliers use the SmsManager API, the TelephonyManager API, and the SpeechRecognizer API. 
In the ”Data collection” cluster, the use of the Bluetooth API is common: outliers do not use this API.
Finally, in the ”\covid self-diagnosis” cluster, the outliers use the MediaPlayer API, which is not used by the rest of the cluster apps.

CHABADA allowed us to identify several \covidRelated apps that deviate from the expected behavior given their description. 
Although the outliers do not necessarily present a danger for end-users, because, in general, they only deviate with respect to non-sensitive APIs (e.g., MediaPlayer, Vibrator, etc.), our empirical results show that descriptions do not always reliably approximate the expected app behavior.

\begin{table}[ht]
    \centering
    \begin{tabular}{c|l|c|c}
    id & Cluster & \# of apps & \# outliers \\ 
    \hline
    0 & Spread tracking & 30 & 3\\
    1 & Sharing health information & 12 & 2 \\
    2 & Sharing general information & 10 & 4 \\
    3 & Data collection & 12 & 4 \\
    4 & \covid self-diagnosis & 14 & 3 \\
    \end{tabular}
    \caption{Clusters of apps generated by our implementation of CHABADA}
    \label{table:clusters}
\end{table}

\highlight{
    \textbf{RQ 6 Answer:}
    Compared to previous studies~\citep{arzt2014flowdroid, kruger2017cognicrypt} on Android apps in general, we note that \covidRelated apps offer a satisfying level of security. Indeed, \covidRelated apps do not appear to leak sensitive data; they are not flagged by AntiVirus engines; Furthermore, they are free of common cryptography API misuses. Nevertheless, several \covidRelated apps have a description that does not match their expected behavior (approximated by their API calls usage). Developers can exploit these empirical results to draw best practices. Similarly, the insights of this study can be leveraged by researchers to develop specific techniques that learn about/from \covidRelated apps.
}

%% file: info.tex
\begin{tikzpicture}
    \tikzstyle{arrowStyle}=[-latex]
    \tikzset{node/.style={minimum height=0.9cm, rounded corners, text width=1.5cm,align=center,scale=.8}}
    \path
    (0,0) node[node,draw] (informationbroadcast) {Information Broadcast}
    ++(-3,-1) node[node,draw] (guidelines) {Guidelines}
    ++(3,0) node[node,draw] (infectionStatistics) {Infection Statistics}
    ++(3,0) node[node,draw] (generalaboutcovid) {General}
    ++(-8,-1) node[node,draw] (zonestoavoid) {Zones to avoid}
    ++(4,0) node[node,draw] (behavioral) {Behavioral}
    ++(3,0) node[node,draw] (filtered) {Filtered}
    ++(2,0) node[node,draw] (notfiltered) {Not filtered}
    ++(-10,-1) node[node,draw] (map) {Map}
    ++(2,0) node[node,draw] (textual) {Textual}
    ++(1.5,0) node[node,draw] (written) {Written}
    ++(1.5,0) node[node,draw] (video) {Video}
    ++(1.5,0) node[node,draw] (vocal) {Vocal};

    \node[draw=none,below] at (infectionStatistics.south) {\tiny 22};
    \node[draw=none,below] at (textual.south) {\tiny 2};
    \node[draw=none,below] at (map.south) {\tiny 3};
    \node[draw=none,below] at (written.south) {\tiny 32};
    \node[draw=none,below] at (video.south) {\tiny 1};
    \node[draw=none,below] at (vocal.south) {\tiny 4};
    \node[draw=none,below] at (filtered.south) {\tiny 15};
    \node[draw=none,below] at (notfiltered.south) {\tiny 14};

    \draw[] (informationbroadcast) -- (guidelines);
    \draw[] (informationbroadcast) -- (infectionStatistics);
    \draw[] (informationbroadcast) -- (generalaboutcovid);
    \draw[] (guidelines) -- (zonestoavoid);
    \draw[] (guidelines) -- (behavioral);
    \draw[] (zonestoavoid) -- (map);
    \draw[] (zonestoavoid) -- (textual);
    \draw[] (behavioral) -- (written);
    \draw[] (behavioral) -- (video);
    \draw[] (behavioral) -- (vocal);
    \draw[] (generalaboutcovid) -- (filtered);
    \draw[] (generalaboutcovid) -- (notfiltered);
\end{tikzpicture}

%% file: upstream.tex
\begin{tikzpicture}
    \tikzstyle{arrowStyle}=[-latex]
    \tikzset{node/.style={minimum height=0.9cm, rounded corners, text width=1.5cm,align=center,scale=.8}}
    \path
    (0,0) node[node,draw] (upstream) {Upstream}
    ++(-3,-1) node[node,draw] (datacollection) {Data Collection}
    ++(7,0) node[node,draw] (spreadtracking) {Spread Tracking}
    ++(-9,-1) node[node,draw] (personal) {Personal}
    ++(2,0) node[node,draw] (medicalreport) {Medical Report}
    ++(2,0) node[node,draw] (reportcase) {Report Case}
    ++(3,0) node[node,draw] (geomonitoring) {Geo-Monitoring}
    ++(2,0) node[node,draw] (socialdistancing) {Social-Distancing}
    ++(2,0) node[node,draw] (contacttracing) {Contact Tracing}
    ++(-11,-1) node[node,draw] (areyouinfected) {Are you infected?}
    ++(2,0) node[node,draw] (healthdiary) {Health Diary}
    ++(2,0) node[node,draw] (currenttreatments) {Current Treatments}
    ++(2,0) node[node,draw] (automatic) {Automatic}
    ++(2,0) node[node,draw] (manual) {Manual}
    ++(1.5,0) node[node,draw] (gps) {GPS}
    ++(1.5,0) node[node,draw] (locationdiary) {Location Diary}
    ++(1.5,0) node[node,draw] (bluetooth) {Bluetooth}
    ++(-7.5,-1) node[node,draw] (diagnosis) {Diagnosis}
    ++(2,0) node[node,draw] (qrcode) {QR code Scanning}
    ++(2,0) node[node,draw] (user) {User data}
    ++(-6,-1) node[node,draw] (automated) {Automated}
    ++(4,0) node[node,draw] (professional) {Professional}
    ++(-5,-1) node[node,draw] (questionnaire) {Form}
    ++(2,0) node[node,draw] (virtualassistant) {Virtual Assistant}
    ++(3,0) node[node,draw] (liveassistance) {Live Assistance}
    ++(-2,-1) node[node,draw] (phone) {Phone}
    ++(2,0) node[node,draw] (instantmessage) {Instant Messages}
    ++(2,0) node[node,draw] (video) {Video};

    \node[draw=none,below] at (personal.south) {\tiny 18};
    \node[draw=none,below] at (reportcase.south) {\tiny 4};
    \node[draw=none,below] at (socialdistancing.south) {\tiny 2};
    \node[draw=none,below] at (areyouinfected.south) {\tiny 25};
    \node[draw=none,below] at (healthdiary.south) {\tiny 17};
    \node[draw=none,below] at (currenttreatments.south) {\tiny 2};
    \node[draw=none,below] at (automatic.south) {\tiny 18};
    \node[draw=none,below] at (user.south) {\tiny 2};
    \node[draw=none,below] at (gps.south) {\tiny 12};
    \node[draw=none,below] at (locationdiary.south) {\tiny 6};
    \node[draw=none,below] at (bluetooth.south) {\tiny 13};
    \node[draw=none,below] at (qrcode.south) {\tiny 1};
    \node[draw=none,below] at (questionnaire.south) {\tiny 35};
    \node[draw=none,below] at (virtualassistant.south) {\tiny 3};
    \node[draw=none,below] at (phone.south) {\tiny 5};
    \node[draw=none,below] at (instantmessage.south) {\tiny 2};
    \node[draw=none,below] at (video.south) {\tiny 4};

    \draw[] (upstream) -- (spreadtracking);
    \draw[] (upstream) -- (diagnosis);
    \draw[] (upstream) -- (datacollection);
    \draw[] (spreadtracking) -- (geomonitoring);
    \draw[] (spreadtracking) -- (socialdistancing);
    \draw[] (spreadtracking) -- (contacttracing);
    \draw[] (diagnosis) -- (automated);
    \draw[] (diagnosis) -- (professional);
    \draw[] (datacollection) -- (personal);
    \draw[] (datacollection) -- (medicalreport);
    \draw[] (datacollection) -- (reportcase);
    \draw[] (geomonitoring) -- (automatic);
    \draw[] (geomonitoring) -- (manual);
    \draw[] (contacttracing) -- (gps);
    \draw[] (contacttracing) -- (locationdiary);
    \draw[] (contacttracing) -- (bluetooth);
    \draw[] (automated) -- (questionnaire);
    \draw[] (automated) -- (virtualassistant);
    \draw[] (professional) -- (liveassistance);
    \draw[] (medicalreport) -- (areyouinfected);
    \draw[] (medicalreport) -- (healthdiary);
    \draw[] (medicalreport) -- (currenttreatments);
    \draw[] (manual) -- (qrcode);
    \draw[] (manual) -- (user);
    \draw[] (liveassistance) -- (phone);
    \draw[] (liveassistance) -- (instantmessage);
    \draw[] (liveassistance) -- (video);
\end{tikzpicture}

%% file: tooling.tex
\begin{tikzpicture}
    \tikzstyle{arrowStyle}=[-latex]
    \tikzset{node/.style={minimum height=1.1cm, rounded corners, text width=2cm,align=center,scale=.8}}
    \path
    (0,0) node[node,draw] (tooling) {Tooling}
    ++(-4.4,-1.5) node[node,draw] (documents) {Document generation for authorities}
    ++(2.2,0) node[node,draw] (appointments) {Appointments}
    ++(2.2,0) node[node,draw] (sales) {Sales}
    ++(2.2,0) node[node,draw] (entertainment) {Entertainment}
    ++(2.2,0) node[node,draw] (education) {Education};

    \node[draw=none,below] at (documents.south) {\tiny 4};
    \node[draw=none,below] at (appointments.south) {\tiny 1};
    \node[draw=none,below] at (sales.south) {\tiny 2};
    \node[draw=none,below] at (entertainment.south) {\tiny 1};
    \node[draw=none,below] at (education.south) {\tiny 1};

    \draw[] (tooling) -- (documents.north);
    \draw[] (tooling) -- (appointments.north);
    \draw[] (tooling) -- (sales.north);
    \draw[] (tooling) -- (entertainment.north);
    \draw[] (tooling) -- (education.north);
\end{tikzpicture}

%% file: first_part_characteristics.tex
\begin{adjustbox}{width=\textwidth}
\begin{tabular}{|l|l|l|l|c|c|c|c|c|c|c|c|}
\hline
\rowcolor{gray!50}
    &   &   &   & 	  \rotCell{1}{l|}{90}{General Information Filtered}   & 	  \rotCell{1}{l|}{90}{Info not filtered}   &   \rotCell{1}{l|}{90}{Vocal Guidelines}   &   \rotCell{1}{l|}{90}{Video Guidelines}   &   \rotCell{1}{l|}{90}{Written Guidelines}   &   \rotCell{1}{l|}{90}{Map Zones To Avoid}   &   \rotCell{1}{l|}{90}{Sales}   & 	  \rotCell{1}{l|}{90}{Textual Zones to Avoid}   \\
    \hline
    \rowcolor{gray!50}
    \multicolumn{1}{|l|}{Package name} & \multicolumn{1}{c|}{Country} & \multicolumn{1}{c|}{Developer type} & \multicolumn{1}{c|}{Target} & \multicolumn{8}{c|}{Information Broadcast} \\
    \hline 
com.gov.mcmc.projectcatur   &   Malaysia   &   Governmental   &   Citizen   & 	  \checkmark   & 	     &      &      &   \checkmark   &      &      & 	     \\
\rowcolor{gray!20}
    com.mohw.corona   &   South Korea   &   Governmental   &   Foreigners   & 	     & 	     &      &      &      &      &      & 	     \\
kg.cdt.stopcovid19   &   Kyrgyztan   &   Governmental   &   Citizen   & 	     & 	     &      &      &      &      &      & 	     \\
\rowcolor{gray!20}
pl.nask.mobywatel   &   Poland   &   Governmental   &   Citizen   & 	  \checkmark   & 	     &      &      &   \checkmark   &      &      & 	     \\
app.ceylon.selftrackingapp   &   Sri Lanka   &   Governmental   &   Citizen   & 	  \checkmark   & 	     &      &      &      &      &      & 	     \\
\rowcolor{gray!20}
bf.diagnoseme.fasocivic   &   Burkina Faso   &   Association   &   Citizen   & 	     & 	  \checkmark   &   \checkmark   &   \checkmark   &   \checkmark   &   \checkmark   &   \checkmark   & 	     \\
com.moc.gh   &   Ghana   &   Governmental   &   Citizen   & 	     & 	     &      &      &      &   \checkmark   &      & 	     \\
\rowcolor{gray!20}
com.vost.covid19mobile   &   Portugal   &   Governmental   &   Citizen   & 	     & 	  \checkmark   &      &      &   \checkmark   &      &      & 	     \\
sg.gov.tech.bluetrace   &   Singapore   &   Governmental   &   Citizen   & 	     & 	     &      &      &   \checkmark   &      &      & 	     \\
\rowcolor{gray!20}
com.joinzoe.covid\_zoe   &   UK/Sweden   &   Company   &   Citizen   & 	     & 	     &      &      &      &      &      & 	     \\
am.gov.covid19   &   Armenia   &   Governmental   &   Citizen   & 	     & 	     &      &      &      &      &      & 	     \\
\rowcolor{gray!20}
sa.gov.nic.tawakkalna   &   Saudi Arabia   &   Governmental   &   Citizen   & 	     & 	     &      &      &      &      &      & 	  \checkmark   \\
fr.gouv.android.stopcovid   &   France   &   Governmental   &   Citizen   & 	     & 	     &      &      &   \checkmark   &      &      & 	     \\
\rowcolor{gray!20}
is.landlaeknir.rakning   &   Iceland   &   Governmental   &   Citizen   & 	     & 	     &      &      &      &      &      & 	     \\
jo.gov.moh.aman   &   Jordan   &   Governmental   &   Citizen   & 	     & 	     &      &      &   \checkmark   &      &      & 	     \\
\rowcolor{gray!20}
nz.govt.health.covidtracer   &   New Zealand   &   Governmental   &   Citizen   & 	     & 	     &      &      &      &      &      & 	     \\
es.gob.asistenciacovid19   &   Spain   &   Governmental   &   Citizen   & 	     & 	  \checkmark   &      &      &      &      &      & 	     \\
\rowcolor{gray!20}
com.moi.covid19   &   Qatar   &   Governmental   &   Citizen   & 	     & 	  \checkmark   &      &      &      &      &      & 	     \\
au.gov.health.covid19   &   Australia   &   Governmental   &   Citizen   & 	  \checkmark   & 	     &      &      &   \checkmark   &      &      & 	     \\
\rowcolor{gray!20}
at.roteskreuz.stopcorona   &   Autriche   &   NGO   &   Citizen   & 	     & 	     &      &      &      &      &      & 	     \\
com.proudcrowd.care   &   USA   &   Company   &   Citizen   & 	     & 	     &      &      &      &      &      & 	     \\
\rowcolor{gray!20}
com.ri.crushcovid   &   USA   &   Governmental   &   Citizen   & 	     & 	  \checkmark   &      &      &      &      &      & 	     \\
au.gov.health.covidsafe   &   Australia   &   Governmental   &   Citizen   & 	     & 	     &      &      &   \checkmark   &      &      & 	     \\
\rowcolor{gray!20}
ca.bc.gov.health.hlbc.COVID19   &   Canada   &   Governmental   &   Citizen   & 	  \checkmark   & 	     &      &      &   \checkmark   &      &      & 	     \\
de.kreativzirkel.coronika   &   Germany   &   Company   &   Citizen   & 	     & 	     &      &      &   \checkmark   &      &      & 	     \\
\rowcolor{gray!20}
    \multicolumn{1}{|p{52mm}|}{\cellcolor{gray!20}com.coronacheck.haugxhaug \newline .testyourcorona}   &   \multirow{2}{*}{Germany}   &   \multirow{2}{*}{Researchers}   &   \multirow{2}{*}{Citizen}   & 	  \multirow{2}{*}{\checkmark}   & 	     &      &      &   \multirow{2}{*}{\checkmark}   &      &      & 	     \\
cat.gencat.mobi.StopCovid19Cat   &   Spain   &   Governmental   &   Citizen   & 	     & 	     &      &      &   \checkmark   &      &      & 	     \\
\rowcolor{gray!20}
cat.gencat.mobi.confinApp   &   Spain   &   Governmental   &   Citizen   & 	  \checkmark   & 	     &      &      &      &      &      & 	     \\
com.ambulis.aphm.covid   &   France   &   Company   &   Patients   & 	     & 	     &      &      &   \checkmark   &      &      & 	     \\
\rowcolor{gray!20}
fr.aphp.covidom   &   France   &   Hospital   &   Patients   & 	     & 	     &      &      &      &      &      & 	     \\
    \multicolumn{1}{|p{18mm}|}{\cellcolor{white!20}appinventor.ai\_david\_taylor\newline.Coronavirus\_help2020}   &   \multirow{2}{*}{Worldwide}   &   \multirow{2}{*}{Independent}   &   \multirow{2}{*}{Citizen}   & 	     & 	  \multirow{2}{*}{\checkmark}   &      &      &      &      &      & 	     \\
\rowcolor{gray!20}
uy.gub.salud.plancovid19uy   &   Uruguay   &   Governmental   &   Citizen   & 	     & 	     &      &      &      &      &      & 	     \\
com.covid19\_algeria   &   Algeria   &   Governmental   &   Citizen   & 	     & 	  \checkmark   &      &      &      &      &      & 	     \\
\rowcolor{gray!20}
com.agetic.coronavirusapp   &   Bolivia   &   Governmental   &   Citizen   & 	  \checkmark   & 	     &      &      &   \checkmark   &      &      & 	     \\
ch.covid19bs.app.PMSMobile   &   Switzerland   &   Governmental   &   Citizen   & 	     & 	     &      &      &   \checkmark   &      &      & 	     \\
\rowcolor{gray!20}
es.gva.coronavirus   &   Spain   &   Governmental   &   Citizen   & 	     & 	  \checkmark   &      &      &      &      &      & 	     \\
    com.govpk.covid19   &   Pakistan   &   Governmental   &   Citizen   & 	     & 	     &      &      &      &      &      & 	     \\
\rowcolor{gray!20}
    \multicolumn{1}{|p{52mm}|}{\cellcolor{gray!20}covid19care.virus.coronavirus\newline.corona.sick.marcom.health.pakistan}   &   \multirow{2}{*}{Pakistan}   &   \multirow{2}{*}{Company}   &   \multicolumn{1}{p{15mm}|}{\cellcolor{gray!20}Media \newline Journalists}   & 	     & 	  \multirow{2}{*}{\checkmark}   &      &      &   \multirow{2}{*}{\checkmark}   &      &      & 	     \\
    \multicolumn{1}{|p{18mm}|}{\cellcolor{white!20}com.tommasomauriello\newline.autocertificazionecoronavirus}   &   \multirow{2}{*}{Italia}   &   \multirow{2}{*}{Independent}   &   \multirow{2}{*}{Citizen}   & 	     & 	     &      &      &      &      &      & 	     \\
\rowcolor{gray!20}
ca.gc.hcsc.canada.covid19   &   Canada   &   Governmental   &   Citizen   & 	  \checkmark   & 	  \checkmark   &      &      &   \checkmark   &      &      & 	     \\
com.krrsoftwaresolutions11.Facemasks   &   Undefined   &   Company   &   Consumers   & 	     & 	     &      &      &      &      &   \checkmark   & 	     \\
\rowcolor{gray!20}
it.adilife.covid19.app   &   Italia   &   Company   &   Citizen   & 	     & 	     &      &      &      &      &      & 	     \\
it.adl.aslroma3.covid19.app   &   Italia   &   Company   &   Citizen   & 	     & 	     &      &      &      &      &      & 	     \\
\rowcolor{gray!20}
com.coronavirus.facemask   &   Undefined   &   Company   &   Users   & 	     & 	     &      &      &      &      &      & 	     \\
com.osapps.covid19   &   Undefined   &   Company   &   Citizen   & 	     & 	  \checkmark   &      &      &   \checkmark   &      &      & 	     \\
\rowcolor{gray!20}
com.maithu.transplantbuddy.covid19   &   Ireland   &   Company   &   Patients   & 	     & 	     &      &      &      &      &      & 	     \\
ar.gob.coronavirus   &   Argentina   &   Governmental   &   Citizen   & 	     & 	     &      &      &      &      &      & 	     \\
\rowcolor{gray!20}
com.bloomreality.sodi   &   Undefined   &   Company   &   Citizen   & 	     & 	     &      &      &      &      &      & 	     \\
br.gov.datasus.guardioes   &   Brazil   &   Governmental   &   Citizen   & 	  \checkmark   & 	     &      &      &   \checkmark   &   \checkmark   &      & 	     \\
\rowcolor{gray!20}
cz.covid19cz.erouska   &   Czech Republic   &   Governmental   &   Citizen   & 	     & 	     &      &      &   \checkmark   &      &      & 	     \\
com.covid19.dgmup   &   India   &   Governmental   &   Citizen   & 	     & 	     &      &      &   \checkmark   &      &      & 	     \\
\rowcolor{gray!20}
com.docandu.checker   &   Greece   &   Company   &   Citizen   & 	     & 	     &      &      &  \checkmark    &      &      & 	     \\
\multicolumn{1}{|p{22mm}|}{\cellcolor{white!20}it.softmining.projects.covid19 \newline .savelifestyle}   &   \multirow{2}{*}{Italia}   &   \multirow{2}{*}{Company}   &   \multirow{2}{*}{Citizen}   & 	     & 	     &   \multirow{2}{*}{\checkmark}   &      &      &      &      & 	     \\
\rowcolor{gray!20}
nic.goi.aarogyasetu   &   India   &   Governmental   &   Citizen   & 	     & 	     &      &      &      &      &      & 	  \checkmark   \\
com.dawsoftware.contacttracker   &   Undefined   &   Company   &   Citizen   & 	     & 	     &      &      &      &      &      & 	     \\
\rowcolor{gray!20}
com.hamagen   &   Israel   &   Governmental   &   Citizen   & 	     & 	     &      &      &      &      &      & 	     \\
hu.gov.virusradar   &   Hungary   &   Governmental   &   Citizen   & 	     & 	     &      &      &      &      &      & 	     \\
\rowcolor{gray!20}
world.coalition.app   &   Undefined   &   Company   &   Citizen   & 	     & 	     &      &      &      &      &      & 	     \\
mk.gov.koronavirus.stop   &   Macedonian   &   Governmental   &   Citizen   & 	     & 	     &      &      &      &      &      & 	     \\
\rowcolor{gray!20}
org.sshield.selfshield   &  Common \newline Weath   &   Association   &   Citizen   & 	     & 	     &      &      &      &      &      & 	     \\
org.pathcheck.covidsafepaths   &   Undefined   &   Association   &   Citizen   & 	     & 	     &      &      &      &      &      & 	     \\
\rowcolor{gray!20}
\multirow{2}{*}{com.knasirayaz.mohapcovid}   &   \multicolumn{1}{p{22mm}|}{\cellcolor{gray!20}United Arab \newline Emirates}   &   \multirow{2}{*}{Governmental}   &   \multirow{2}{*}{Citizen}   & 	  \multirow{2}{*}{\checkmark}   & 	     &      &      &   \checkmark   &      &      & 	     \\
gov.georgia.novid20   &   Georgia   &   Governmental   &   Citizen   & 	     & 	     &      &      &      &      &      & 	     \\
\rowcolor{gray!20}
nl.lumc.covidradar   &   Netherlands   &   Governmental   &   Citizen   & 	     & 	  \checkmark   &      &      &   \checkmark   &      &      & 	     \\
com.Eha.covid\_19   &   Vietnam   &   Governmental   &   Citizen   & 	     & 	  \checkmark   &   \checkmark   &      &   \checkmark   &      &      & 	     \\
\rowcolor{gray!20}
covid.trace.morocco   &   Morocco   &   Governmental   &   Citizen   & 	     & 	     &      &      &   \checkmark   &      &      & 	     \\
ru.mos.socmon   &   Russia   &   Governmental   &   Patients   & 	     & 	     &      &      &   \checkmark   &      &      & 	     \\
\rowcolor{gray!20}
org.prixa.p5covidtracker   &   Nepal   &   Governmental   &   Citizen   & 	  \checkmark   & 	     &      &      &      &      &      & 	     \\
de.bssd.covid19   &   Germany   &   Company   &   Patients   & 	     & 	     &      &      &      &      &      & 	     \\
\rowcolor{gray!20}
com.edu.aku.akuhccheck   &   Pakistan   &   Researchers   &   Citizen   & 	  \checkmark   & 	     &   \checkmark   &      &   \checkmark   &      &      & 	     \\
\multirow{2}{*}{com.pixxonai.covid19}   &   \multirow{2}{*}{India}   &   \multirow{2}{*}{Governmental}   &   \multicolumn{1}{p{5mm}|}{\cellcolor{white!20}Quarantine \newline persons}   & 	     & 	     &      &      &      &      &      & 	     \\
\rowcolor{gray!20}
de.rki.coronadatenspende   &   Germany   &   Governmental   &   Citizen   & 	     & 	     &      &      &      &      &      & 	     \\
mx.gob.www   &   Mexico   &   Governmental   &   Citizen   & 	  \checkmark   & 	     &      &      &   \checkmark   &      &      & 	     \\
\rowcolor{gray!20}
tr.gov.saglik.koronaonlem   &   Turkey   &   Governmental   &   Citizen   & 	     & 	     &      &      &   \checkmark   &      &      & 	     \\
com.telkom.tracencare   &   Indonesia   &   Governmental   &   Citizen   & 	     & 	     &      &      &      &      &      & 	     \\
\rowcolor{gray!20}
co.gov.ins.guardianes   &   Colombia   &   Governmental   &   Citizen   & 	  \checkmark   & 	     &      &      &      &      &      & 	     \\
bh.bahrain.corona.tracker   &   Bahrain   &   Governmental   &   Citizen   & 	     & 	     &      &      &   \checkmark   &      &      & 	     \\
\rowcolor{gray!20}
bg.government.virusafe   &   Bulgaria   &   Governmental   &   Citizen   & 	     & 	  \checkmark   &      &      &      &      &      & 	     \\
\hline
\end{tabular}
\end{adjustbox}

%% file: second_part_characteristics.tex
\begin{adjustbox}{width=\textwidth}
    {\scriptsize
\begin{tabular}{|l|l|l|l|c|c|c|c|c|c|c|c|}
\hline
\rowcolor{gray!50}
      &  	  \rotCell{1}{l|}{90}{Health Diary}    &    \rotCell{1}{l|}{90}{Qr Code Tracing}    &    \rotCell{1}{l|}{90}{Automated Geo-Monitoring}    &    \rotCell{1}{l|}{90}{Are You Infected?}    &    \rotCell{1}{l|}{90}{Report Covid Case}    &    \rotCell{1}{l|}{90}{GPS Contact Tracing}    &  	  \rotCell{1}{l|}{90}{Automatic Diagnosis}   &    \rotCell{1}{l|}{90}{Live-Assistance Chat}    &    \rotCell{1}{l|}{90}{Bluetooth Contact Tracing}    &  	  \rotCell{1}{l|}{90}{Personnal Data Collection}    &    \rotCell{1}{l|}{90}{Current Treatment}    \\ 
    \hline
    \rowcolor{gray!50}
    \multicolumn{1}{|l|}{Package name} & \multicolumn{11}{c|}{Upstream} \\
    \hline 
com.gov.mcmc.projectcatur     &  	      &        &        &        &        &    \checkmark    &  	      &        &        &  	      &        \\ 
    \rowcolor{gray!20}
    com.mohw.corona     &  	  \checkmark    &        &        &        &        &        &  	      &        &        &  	      &        \\ 
kg.cdt.stopcovid19     &  	      &        &    \checkmark    &    \checkmark    &    \checkmark    &        &  	      &        &        &  	      &        \\ 
    \rowcolor{gray!20}
pl.nask.mobywatel     &  	      &        &        &        &        &         &  	      &        &        &  	      &        \\ 
app.ceylon.selftrackingapp     &  	      &        &    \checkmark    &        &        &        &  	      &        &        &  	      &        \\ 
    \rowcolor{gray!20}
bf.diagnoseme.fasocivic     &  	      &        &        &        &        &        &  	  \checkmark    &    \checkmark    &        &  	      &        \\ 
com.moc.gh     &  	      &        &        &        &        &        &  	  \checkmark    &        &        &  	      &        \\ 
    \rowcolor{gray!20}
com.vost.covid19mobile     &  	      &        &        &        &        &        &  	      &        &        &  	      &        \\ 
sg.gov.tech.bluetrace     &  	      &        &        &        &        &        &  	      &        &    \checkmark    &  	      &        \\ 
    \rowcolor{gray!20}
com.joinzoe.covid\_zoe     &  	  \checkmark    &        &        &    \checkmark    &        &        &  	  \checkmark    &        &        &  	  \checkmark    &    \checkmark    \\ 
am.gov.covid19     &  	      &        &        &        &        &        &  	  \checkmark    &        &        &  	      &        \\ 
    \rowcolor{gray!20}
sa.gov.nic.tawakkalna     &  	      &        &        &        &    \checkmark    &        &  	  \checkmark    &        &        &  	      &        \\ 
fr.gouv.android.stopcovid     &  	      &        &        &    \checkmark    &        &        &  	      &        &    \checkmark    &  	      &        \\ 
    \rowcolor{gray!20}
is.landlaeknir.rakning     &  	      &        &    \checkmark    &    \checkmark    &        &    \checkmark    &  	      &        &        &  	      &        \\ 
jo.gov.moh.aman     &  	      &        &    \checkmark    &    \checkmark    &        &        &  	      &        &        &  	      &        \\ 
    \rowcolor{gray!20}
nz.govt.health.covidtracer     &  	      &    \checkmark     &        &        &        &        &  	      &        &        &  	  \checkmark    &        \\ 
es.gob.asistenciacovid19     &  	      &        &    \checkmark    &        &        &        &  	  \checkmark    &        &        &  	  \checkmark    &        \\ 
    \rowcolor{gray!20}
com.moi.covid19     &  	      &        &    \checkmark    &        &        &        &  	      &        &        &  	  \checkmark    &        \\ 
au.gov.health.covid19     &  	      &        &        &        &        &        &  	  \checkmark    &        &        &  	      &        \\ 
    \rowcolor{gray!20}
at.roteskreuz.stopcorona     &  	      &        &        &    \checkmark    &        &       &  	  \checkmark    &        &    \checkmark    &  	      &        \\ 
com.proudcrowd.care     &  	      &        &        &    \checkmark    &        &    \checkmark    &  	      &        &        &  	      &        \\ 
    \rowcolor{gray!20}
com.ri.crushcovid     &  	      &        &    \checkmark    &    \checkmark    &        &    \checkmark    &  	  \checkmark    &        &        &  	      &        \\ 
au.gov.health.covidsafe     &  	      &        &        &    \checkmark    &    \checkmark    &        &  	      &        &    \checkmark    &  	      &        \\ 
    \rowcolor{gray!20}
ca.bc.gov.health.hlbc.COVID19     &  	      &        &        &        &        &    \checkmark    &  	      &        &        &  	  \checkmark    &        \\ 
de.kreativzirkel.coronika     &  	  \checkmark    &        &        &        &        &        &  	      &        &        &  	      &        \\ 
    \rowcolor{gray!20}
    \multicolumn{1}{|p{52mm}|}{\cellcolor{gray!20}com.coronacheck.haugxhaug \newline .testyourcorona}     &  	      &        &        &        &        &        &  	  \multirow{2}{*}{\checkmark}    &        &        &  	      &        \\ 
cat.gencat.mobi.StopCovid19Cat     &  	  \checkmark    &        &        &    \checkmark    &        &        &  	  \checkmark    &        &        &  	      &        \\ 
    \rowcolor{gray!20}
cat.gencat.mobi.confinApp     &  	      &        &        &        &        &        &  	      &        &        &  	      &        \\ 
com.ambulis.aphm.covid     &  	  \checkmark    &        &        &        &        &        &  	  \checkmark    &        &        &  	      &        \\ 
    \rowcolor{gray!20}
fr.aphp.covidom     &  	  \checkmark    &        &        &        &        &        &  	  \checkmark    &        &        &  	      &        \\ 
    \multicolumn{1}{|p{18mm}|}{\cellcolor{white!20}appinventor.ai\_david\_taylor\newline.Coronavirus\_help2020}     &  	      &        &        &        &        &        &  	      &        &        &  	      &        \\ 
    \rowcolor{gray!20}
uy.gub.salud.plancovid19uy     &  	  \checkmark    &        &        &        &        &        &  	  \checkmark    &        &        &  	  \checkmark    &        \\ 
com.covid19\_algeria     &  	      &        &    \checkmark    &        &        &        &  	      &        &        &  	  \checkmark    &        \\ 
    \rowcolor{gray!20}
com.agetic.coronavirusapp     &  	      &        &        &        &        &        &  	  \checkmark    &        &        &  	      &        \\ 
ch.covid19bs.app.PMSMobile     &  	      &        &        &    \checkmark    &        &        &  	  \checkmark    &        &        &  	      &        \\ 
    \rowcolor{gray!20}
es.gva.coronavirus     &  	      &        &        &        &        &        &  	  \checkmark    &        &        &  	      &        \\ 
    com.govpk.covid19     &  	      &        &        &        &        &        &  	      &        &        &  	      &        \\ 
    \rowcolor{gray!20}
    \multicolumn{1}{|p{52mm}|}{\cellcolor{gray!20}covid19care.virus.coronavirus\newline.corona.sick.marcom.health.pakistan}     &  	      &        &        &        &    \multirow{2}{*}{\checkmark}    &        &  	      &        &        &  	      &        \\ 
    \multicolumn{1}{|p{18mm}|}{\cellcolor{white!20}com.tommasomauriello\newline.autocertificazionecoronavirus}     &  	      &        &        &        &        &        &  	      &        &        &  	  \multirow{2}{*}{\checkmark}    &        \\ 
    \rowcolor{gray!20}
ca.gc.hcsc.canada.covid19     &  	      &        &        &        &        &        &  	  \checkmark    &        &        &  	      &        \\ 
com.krrsoftwaresolutions11.Facemasks     &  	      &        &        &        &        &        &  	      &        &        &  	      &        \\ 
    \rowcolor{gray!20}
it.adilife.covid19.app     &  	  \checkmark    &        &        &        &        &       &  	  \checkmark    &        &        &  	  \checkmark    &        \\ 
it.adl.aslroma3.covid19.app     &  	  \checkmark    &        &        &        &        &        &  	  \checkmark    &        &        &  	  \checkmark    &        \\ 
    \rowcolor{gray!20}
com.coronavirus.facemask     &  	      &        &        &        &        &        &  	      &        &        &  	      &        \\ 
com.osapps.covid19     &  	      &        &        &        &        &        &  	      &        &        &  	      &        \\ 
    \rowcolor{gray!20}
com.maithu.transplantbuddy.covid19     &  	  \checkmark    &        &    \checkmark    &        &        &        &  	  \checkmark    &        &        &  	      &    \checkmark    \\ 
ar.gob.coronavirus     &  	      &        &    \checkmark    &        &        &        &  	  \checkmark    &        &        &  	      &        \\ 
    \rowcolor{gray!20}
com.bloomreality.sodi     &  	      &        &        &        &        &        &  	      &        &        &  	      &        \\ 
br.gov.datasus.guardioes     &  	      &        &        &        &        &        &  	  \checkmark    &        &        &  	      &        \\ 
    \rowcolor{gray!20}
cz.covid19cz.erouska     &  	      &        &        &    \checkmark    &        &        &  	      &        &    \checkmark    &  	      &        \\ 
com.covid19.dgmup     &  	      &        &        &        &        &        &  	      &        &        &  	  \checkmark    &        \\ 
    \rowcolor{gray!20}
com.docandu.checker     &  	      &        &        &        &        &        &  	  \checkmark    &    \checkmark    &        &  	      &        \\ 
\multicolumn{1}{|p{22mm}|}{\cellcolor{white!20}it.softmining.projects.covid19 \newline .savelifestyle}     &  	      &        &        &        &        &    \multirow{2}{*}{\checkmark}    &  	  \multirow{2}{*}{\checkmark}    &        &        &  	      &        \\ 
    \rowcolor{gray!20}
nic.goi.aarogyasetu     &  	      &        &        &    \checkmark    &        &        &  	      &        &    \checkmark    &  	      &        \\ 
com.dawsoftware.contacttracker     &  	  \checkmark    &        &        &        &        &    \checkmark    &  	  \checkmark    &        &        &  	      &        \\ 
    \rowcolor{gray!20}
com.hamagen     &  	      &        &    \checkmark    &    \checkmark    &        &        &  	  \checkmark    &        &        &  	      &        \\ 
hu.gov.virusradar     &  	      &        &        &        &        &    \checkmark    &  	      &        &    \checkmark    &  	      &        \\ 
    \rowcolor{gray!20}
world.coalition.app     &  	      &        &        &    \checkmark    &        &        &  	      &        &    \checkmark    &  	      &        \\ 
mk.gov.koronavirus.stop     &  	      &        &        &    \checkmark    &        &        &  	      &        &    \checkmark    &  	      &        \\ 
    \rowcolor{gray!20}
org.sshield.selfshield     &  	  \checkmark    &        &        &        &        &        &  	  \checkmark    &        &        &  	      &        \\ 
org.pathcheck.covidsafepaths     &  	      &        &        &    \checkmark    &        &    \checkmark    &  	      &        &        &  	      &        \\ 
    \rowcolor{gray!20}
com.knasirayaz.mohapcovid     &  	      &        &        &        &        &        &  	      &        &        &  	      &        \\ 
gov.georgia.novid20     &  	      &        &        &    \checkmark    &        &    \checkmark    &  	      &        &    \checkmark    &  	      &        \\ 
    \rowcolor{gray!20}
nl.lumc.covidradar     &  	      &        &        &        &        &        &  	      &        &        &  	  \checkmark    &        \\ 
com.Eha.covid\_19     &  	      &        &    \checkmark    &        &        &        &  	      &        &        &  	      &        \\ 
    \rowcolor{gray!20}
covid.trace.morocco     &  	      &        &        &    \checkmark    &        &        &  	      &        &    \checkmark    &  	      &        \\ 
ru.mos.socmon     &  	      &        &    \checkmark    &        &        &        &  	      &        &        &  	  \checkmark    &        \\ 
    \rowcolor{gray!20}
org.prixa.p5covidtracker     &  	  \checkmark    &        &        &    \checkmark    &        &        &  	  \checkmark    &        &        &  	  \checkmark    &        \\ 
de.bssd.covid19     &  	      &        &        &    \checkmark    &        &        &  	      &        &        &  	      &        \\ 
    \rowcolor{gray!20}
com.edu.aku.akuhccheck     &  	      &        &        &        &        &        &  	  \checkmark    &        &        &  	      &        \\ 
com.pixxonai.covid19     &  	  \checkmark    &        &    \checkmark    &        &        &        &  	  \checkmark    &        &        &  	  \checkmark    &        \\ 
    \rowcolor{gray!20}
de.rki.coronadatenspende     &  	  \checkmark    &        &        &        &        &        &  	      &        &        &  	      &        \\ 
mx.gob.www     &  	      &        &        &        &        &        &  	  \checkmark    &        &        &  	      &        \\ 
    \rowcolor{gray!20}
tr.gov.saglik.koronaonlem     &  	      &        &    \checkmark    &        &        &        &  	  \checkmark    &        &        &  	  \checkmark    &        \\ 
com.telkom.tracencare     &  	      &        &        &    \checkmark    &        &        &  	      &        &    \checkmark    &  	      &        \\ 
    \rowcolor{gray!20}
co.gov.ins.guardianes     &  	  \checkmark    &        &    \checkmark    &    \checkmark    &        &        &  	  \checkmark    &        &        &  	  \checkmark    &        \\ 
bh.bahrain.corona.tracker     &  	      &        &    \checkmark    &    \checkmark    &        &    \checkmark    &  	      &        &    \checkmark    &  	  \checkmark    &        \\ 
    \rowcolor{gray!20}
bg.government.virusafe     &  	  \checkmark    &        &    \checkmark    &    \checkmark    &        &    \checkmark    &  	  \checkmark    &        &        &  	      &        \\ 
\hline
\end{tabular}
}
\end{adjustbox}

%% file: third_part_characteristics.tex
\begin{adjustbox}{width=\textwidth}
    {\scriptsize
\begin{tabular}{|l|l|l|l|c|c|c|c|c|c|c|c|}
\hline
\rowcolor{gray!50}
    & 	   \rotCell{1}{l|}{90}{Live-Assistance Phone}    &    \rotCell{1}{l|}{90}{User data Geo-Monitoring}    &    \rotCell{1}{l|}{90}{Location Diary}    &    \rotCell{1}{l|}{90}{Virtual Assistant}    &    \rotCell{1}{l|}{90}{Live-Assistance Video}    &    \rotCell{1}{l|}{90}{Social-Distancing}  & 	  \rotCell{1}{l|}{90}{Appointment}   &   \rotCell{1}{l|}{90}{Entertainment}   & 	  \rotCell{1}{l|}{90}{Sales}   &   \rotCell{1}{l|}{90}{Education}   &   \rotCell{1}{l|}{90}{Document Creation}  \\
    \hline
    \rowcolor{gray!50}
    \multicolumn{1}{|l|}{Package name} & \multicolumn{6}{c|}{Upstream} & \multicolumn{5}{c|}{Tooling}\\
    \hline
com.gov.mcmc.projectcatur    & 	       &        &        &        &        &      & 	     &      & 	     &      &     \\
\rowcolor{gray!20}
com.mohw.corona    & 	       &        &        &        &        &      & 	     &      & 	     &      &     \\
kg.cdt.stopcovid19    & 	       &        &        &        &        &      & 	     &      & 	     &      &     \\
\rowcolor{gray!20}
pl.nask.mobywatel    & 	       &        &        &        &        &      & 	     &      & 	     &      &     \\
app.ceylon.selftrackingapp    & 	       &        &        &        &        &      & 	     &      & 	     &      &     \\
\rowcolor{gray!20}
bf.diagnoseme.fasocivic    & 	       &        &        &        &        &      & 	     &      & 	  \checkmark   &   \checkmark   &     \\
com.moc.gh    & 	       &        &        &        &        &      & 	     &      & 	     &      &     \\
\rowcolor{gray!20}
com.vost.covid19mobile    & 	       &        &        &        &        &      & 	     &      & 	     &      &     \\
sg.gov.tech.bluetrace    & 	       &        &        &        &        &      & 	     &      & 	     &      &     \\
\rowcolor{gray!20}
com.joinzoe.covid\_zoe    & 	       &        &        &        &        &      & 	     &      & 	     &      &     \\
am.gov.covid19    & 	       &        &        &        &        &      & 	     &      & 	     &      &     \\
\rowcolor{gray!20}
sa.gov.nic.tawakkalna    & 	       &        &        &        &        &      & 	     &      & 	     &      &   \checkmark  \\
fr.gouv.android.stopcovid    & 	       &        &        &        &        &      & 	     &      & 	     &      &     \\
\rowcolor{gray!20}
is.landlaeknir.rakning    & 	       &        &        &        &        &      & 	     &      & 	     &      &     \\
jo.gov.moh.aman    & 	       &        &        &        &        &      & 	     &      & 	     &      &     \\
\rowcolor{gray!20}
nz.govt.health.covidtracer    & 	       &        &    \checkmark    &        &        &      & 	     &      & 	     &      &     \\
es.gob.asistenciacovid19    & 	       &        &        &        &        &      & 	     &      & 	     &      &     \\
\rowcolor{gray!20}
com.moi.covid19    & 	   \checkmark    &        &        &        &        &      & 	     &      & 	     &      &     \\
au.gov.health.covid19    & 	       &        &        &        &        &      & 	     &      & 	     &      &     \\
\rowcolor{gray!20}
at.roteskreuz.stopcorona    & 	       &        &        &        &        &      & 	     &      & 	     &      &     \\
com.proudcrowd.care    & 	       &    \checkmark    &    \checkmark    &        &        &      & 	     &      & 	     &      &     \\
\rowcolor{gray!20}
com.ri.crushcovid    & 	       &        &    \checkmark    &        &        &      & 	     &      & 	     &      &     \\
au.gov.health.covidsafe    & 	       &        &        &        &        &      & 	     &      & 	     &      &     \\
\rowcolor{gray!20}
ca.bc.gov.health.hlbc.COVID19    & 	       &        &        &        &        &      & 	     &      & 	     &      &     \\
de.kreativzirkel.coronika    & 	       &    \checkmark    &        &        &        &      & 	     &      & 	     &      &     \\
\rowcolor{gray!20}
\multicolumn{1}{|p{52mm}|}{\cellcolor{gray!20}com.coronacheck.haugxhaug \newline .testyourcorona}    & 	       &        &        &        &        &      & 	     &      & 	     &      &     \\
cat.gencat.mobi.StopCovid19Cat    & 	       &        &        &        &        &      & 	     &      & 	     &      &     \\
\rowcolor{gray!20}
cat.gencat.mobi.confinApp    & 	       &        &        &    \checkmark    &        &      & 	     &      & 	     &      &   \checkmark  \\
com.ambulis.aphm.covid    & 	       &        &        &        &        &      & 	     &      & 	     &      &     \\
\rowcolor{gray!20}
fr.aphp.covidom    & 	   \checkmark    &        &        &        &        &      & 	     &      & 	     &      &     \\
\multicolumn{1}{|p{18mm}|}{\cellcolor{white!20}appinventor.ai\_david\_taylor\newline.Coronavirus\_help2020}    & 	       &        &        &        &        &      & 	     &      & 	     &      &     \\
\rowcolor{gray!20}
uy.gub.salud.plancovid19uy    & 	       &        &        &        &    \checkmark    &      & 	     &      & 	     &      &     \\
com.covid19\_algeria    & 	       &        &        &        &        &      & 	     &      & 	     &      &     \\
\rowcolor{gray!20}
com.agetic.coronavirusapp    & 	       &        &        &        &        &      & 	     &      & 	     &      &     \\
ch.covid19bs.app.PMSMobile    & 	       &        &        &        &        &      & 	     &      & 	     &      &     \\
\rowcolor{gray!20}
es.gva.coronavirus    & 	       &        &        &        &        &      & 	  \checkmark   &      & 	     &      &     \\
com.govpk.covid19    & 	       &        &        &        &        &      & 	     &      & 	     &      &     \\
\rowcolor{gray!20}
\multicolumn{1}{|p{52mm}|}{\cellcolor{gray!20}covid19care.virus.coronavirus\newline.corona.sick.marcom.health.pakistan}    & 	       &        &        &        &        &      & 	     &      & 	     &      &     \\
\multicolumn{1}{|p{18mm}|}{\cellcolor{white!20}com.tommasomauriello\newline.autocertificazionecoronavirus}    & 	       &        &        &        &        &      & 	     &      & 	     &      &   \multirow{2}{*}{\checkmark}  \\
\rowcolor{gray!20}
ca.gc.hcsc.canada.covid19    & 	       &        &        &        &        &      & 	     &      & 	     &      &     \\
com.krrsoftwaresolutions11.Facemasks    & 	       &        &        &        &        &      & 	     &      & 	  \checkmark   &      &     \\
\rowcolor{gray!20}
it.adilife.covid19.app    & 	       &        &        &        &    \checkmark    &      & 	     &      & 	     &      &     \\
it.adl.aslroma3.covid19.app    & 	       &        &        &        &    \checkmark    &      & 	     &      & 	     &      &     \\
\rowcolor{gray!20}
com.coronavirus.facemask    & 	       &        &        &        &        &      & 	     &   \checkmark   & 	     &      &     \\
com.osapps.covid19    & 	       &        &        &        &        &      & 	     &      & 	     &      &     \\
\rowcolor{gray!20}
com.maithu.transplantbuddy.covid19    & 	       &        &        &        &        &      & 	     &      & 	     &      &     \\
ar.gob.coronavirus    & 	       &        &        &        &        &      & 	     &      & 	     &      &   \checkmark  \\
\rowcolor{gray!20}
com.bloomreality.sodi    & 	       &        &        &        &        &    \checkmark  & 	     &      & 	     &      &     \\
br.gov.datasus.guardioes    & 	       &        &        &        &        &      & 	     &      & 	     &      &     \\
\rowcolor{gray!20}
cz.covid19cz.erouska    & 	   \checkmark    &        &        &        &        &      & 	     &      & 	     &      &     \\
com.covid19.dgmup    & 	       &        &    \checkmark    &        &        &      & 	     &      & 	     &      &     \\
\rowcolor{gray!20}
com.docandu.checker    & 	       &        &        &        &        &      & 	     &      & 	     &      &     \\
\multicolumn{1}{|p{22mm}|}{\cellcolor{white!20}it.softmining.projects.covid19 \newline .savelifestyle}    & 	       &        &        &        &        &      & 	     &      & 	     &      &     \\
\rowcolor{gray!20}
nic.goi.aarogyasetu    & 	       &        &        &        &        &      & 	     &      & 	     &      &     \\
com.dawsoftware.contacttracker    & 	       &        &        &        &        &      & 	     &      & 	     &      &     \\
\rowcolor{gray!20}
com.hamagen    & 	       &        &        &        &        &      & 	     &      & 	     &      &     \\
hu.gov.virusradar    & 	       &        &        &        &        &      & 	     &      & 	     &      &     \\
\rowcolor{gray!20}
world.coalition.app    & 	       &        &        &        &        &      & 	     &      & 	     &      &     \\
mk.gov.koronavirus.stop    & 	       &        &        &        &        &      & 	     &      & 	     &      &     \\
\rowcolor{gray!20}
org.sshield.selfshield    & 	       &        &        &        &        &      & 	     &      & 	     &      &     \\
org.pathcheck.covidsafepaths    & 	       &        &    \checkmark    &        &        &      & 	     &      & 	     &      &     \\
\rowcolor{gray!20}
com.knasirayaz.mohapcovid    & 	   \checkmark    &        &        &        &        &      & 	     &      & 	     &      &     \\
gov.georgia.novid20    & 	       &        &        &        &        &      & 	     &      & 	     &      &     \\
\rowcolor{gray!20}
nl.lumc.covidradar    & 	       &        &        &        &        &      & 	     &      & 	     &      &     \\
com.Eha.covid\_19    & 	       &        &        &    \checkmark    &    \checkmark    &      & 	     &      & 	     &      &     \\
\rowcolor{gray!20}
covid.trace.morocco    & 	       &        &        &        &        &      & 	     &      & 	     &      &     \\
ru.mos.socmon    & 	       &        &        &        &        &      & 	     &      & 	     &      &     \\
\rowcolor{gray!20}
org.prixa.p5covidtracker    & 	       &        &        &        &        &      & 	     &      & 	     &      &     \\
de.bssd.covid19    & 	       &        &        &        &        &      & 	     &      & 	     &      &     \\
\rowcolor{gray!20}
com.edu.aku.akuhccheck    & 	       &        &        &    \checkmark    &        &    \checkmark  & 	     &      & 	     &      &     \\
com.pixxonai.covid19    & 	       &        &        &        &        &      & 	     &      & 	     &      &     \\
\rowcolor{gray!20}
de.rki.coronadatenspende    & 	       &        &        &        &        &      & 	     &      & 	     &      &     \\
mx.gob.www    & 	       &        &        &        &        &     & 	     &      & 	     &      &     \\
\rowcolor{gray!20}
tr.gov.saglik.koronaonlem    & 	       &        &    \checkmark    &        &        &      & 	     &      & 	     &      &     \\
com.telkom.tracencare    & 	   \checkmark    &        &        &        &        &      & 	     &      & 	     &      &     \\
\rowcolor{gray!20}
co.gov.ins.guardianes    & 	       &        &        &        &        &      & 	     &      & 	     &      &     \\
bh.bahrain.corona.tracker    & 	       &        &        &        &        &      & 	     &      & 	     &      &     \\
\rowcolor{gray!20}
bg.government.virusafe    & 	       &        &        &        &        &      & 	     &      & 	     &      &     \\
\hline
\end{tabular}
}
\end{adjustbox}

%% file: discussion.tex
\section{Threat to validity}
\label{discussion}

Our study bears a number of threats to validity related to the selection of apps, external factors that may impact our conclusions, and the limitations of the leveraged tools.

\subsection{\covidRelated apps retrieval}
\paragraph{We may have missed some \covidRelated apps.} Our approach relies on simple heuristics to gather applications from 
\az{}, which helped to identify \covidAppsInAz{} unique apps.
Therefore, we cannot guarantee that our app collection of \covidRelated apps is exhaustive. 
However, we leveraged \az{} which is the largest and continuously-updated repository of Android applications available to the 
research community, which we further manually supplemented with other sources. 
Overall, it is unlikely that we missed a number of \covidRelated apps that is large enough 
to significantly invalidate the conclusions presented in this paper. Actually, as we have revealed, we were able to catch in time some apps that were later removed from the market.

\paragraph{The enforcement of Google Policy on market apps may have biased our study on \covidRelated apps.} Our experiments were conducted until early June 2020, and we know that Google 
has been enforcing its policy regarding \covidRelated apps since at least early 
April~\cite{googleupdate}. 
Given our vantage point without insider knowledge, our observations are limited 
to apps that were actually released on \gp{}, and our study is blind to apps that were never let in \gp{}. 
There is thus a possibility that some of our findings are \emph{consequences} of Google's policy, 
and not a characterization of apps that were meant to be on \gp{}. 
Nonetheless, our study reflects closely the landscape of apps that were made available to users.

On the other hand, we note that the removal of apps from the market may have been performed by either market maintainers or by developers themselves. 
Future studies may attempt to investigate closely the reasons of app removal.

\paragraph{Apk date is unreliable.}
In our collected dataset, the earliest date of appearance of \covidRelated apps goes back to September 2017, 
but we relied on the earliest date between the \az{} added date, the first 
submission date to \vt{}, and several other websites (included last update date in \gp{} as discussed in 
Section~\ref{res:dates}).
Since \az{} does not provide the release date of an app (which is different from the added date), it could be the case that the apps existed before. 
Unfortunately, we cannot rely on the dates of the files constituting the apps since 
they can easily be modified~\citep{moonlightbox}.
Fortunately, Google Play mirrors exist (e.g., appbrain.com) and can keep track of different versions of Android apps, even the first release date. We heavily relied on those dates which mitigates the threats to validity.

\subsection{Reliability of the app description}
As discussed in Section~\ref{subsec:curation}, some parts of our study on \covidRelated apps require manual analysis, e.g., to inspect app’s descriptions or app’s content, or to check the output of the automated analysis tools. While we followed a consistent process by carefully verifying what was under study and by replicating this process multiple times, the conclusions are subject to human subjectivity.

\subsection{Reliability of the manual analyses}
As discussed in Section\ref{manual_analyses}, some parts of our study on \covidRelated apps require manual analysis, e.g., to inspect app’s descriptions or app’s content, or to check the output of the automated analysis tools. While we followed a consistent process by carefully verifying what was under study and by replicating this process multiple times, the conclusions are subject to human subjectivity.
Therefore, the results presented (e.g. the taxonomy) may not be as representative as they should be.
Nevertheless, we strongly believe that, from the process we followed, it is unlikely that the results we collected are not far from the truth.

\subsection{Tools limitations}
We leveraged several security scanners to evaluate different security characteristics of apps, and hence inherit their limitations. 
We cannot guarantee that these tools yielded accurate analyses. We mitigated the threats by first ensuring the selection of state of the art tools that are commonly used in the literature, and second by ensuring that we do not overclaim based on their results. 

%% file: related_work.tex
\section{Related Work} \label{related_work}

Several prior works have conducted empirical studies on large sets of Android applications collected from 
app markets. 
\citet{playdrone} collected more than a million apps from \gp{} and uncovered several interesting patterns 
in Android apps and the way they are developed. 
Also in 2014, \citet{appbehavior} collected apps and their associated descriptions and automatically 
verified whether the descriptions actually matched the app behaviors 
while \citet{10.1145/2660267.2660287} checked the descriptions against the permission usages. 
Other works focused on financial apps~\citep{financialapps}, on app maintenance and prices~\citep{longitudinal}, on malware~\citep{zhou2012dissecting}, or on the quality 
of apps descriptions~\citep{appdescriptions}.

To the best of our knowledge, the academic literature has not yet reported on a 
systematic study on \covidRelated Android apps. 
\citet{he2020virus} however have recently discussed the case of coronavirus-themed malicious apps.
In this paper, the authors analyzed 277 malicious applications related to \covid among an initial dataset
of \num{2016} applications. 
In our paper, we did not identify any malicious \covidRelated apps 
which seems to be in contradiction with \citet{he2020virus} who collected 277 malicious \covidRelated apps. 
However, this discrepancy could be explained by at least two reasons:
\begin{itemize}
    \item We used a more selective filtering process to collect \covidRelated apps, notably because we realized that the keywords we initially used (and that \citet{he2020virus} also used) were 
    too broad, i.e., they catch many apps not related to \covid;
    \item Among their initial dataset of \num{2016} apps, only 6 are coming from \gp{}. Unfortunately, the paper does not precise whether the 6 apps from \gp{} are part of their 277 identified malicious apps. Overall, their results show that the vast majority (and probably all, as per our results) of malicious \covidRelated apps are coming from application sources that are outside \gp{}.  
\end{itemize}

Besides, note that in \citet{he2020virus} we do not have the information whether
they considered different versions of a single app or not. Considering multiple versions 
 could drastically lower the number of apps in their dataset.
Their study further shows that there is a correlation between the number of \covid
cases in the world and the number of malicious \covidRelated apps and 
that malicious developers did not repackage existing \covidRelated apps
for releasing malicious apps, contrary to a common 
malware practice~\citep{zhou2012dissecting}. 
\citet{he2020virus} uncovered 34 different malware families
used in malicious \covidRelated apps. 
Their results further suggest that malware
developers do not target specific users but target a wide range of countries. 

Since the emergence of the \covid topic in the media, researchers are investigating 
its effect, not only on the medical front but also on our daily life.
Related to the security perspective of mobile applications users, 
several security researchers and analysts publicly disclosed their findings, usually 
in blog posts or press articles, about the activity 
of malware developers trying to take advantage of the \covid 
crisis~\citep{forbes,buguroo,domaintools,securityboulevard}. 

Similarly, from a privacy perspective, 
a common functionality of \covidRelated apps for fighting against the spread of
\covid is contact tracing. Tracing however carries several concerns with respect to user privacy. 
Indeed, \citet{baumgrtner2020mind}
show that although developers claim to respect privacy,  it is possible to
de-anonymize information about infected persons that are traced and even sabotage the tracing effort by
injecting fake contacts.

%% file: conclusion.tex
\section{Conclusion} \label{conclusion}

In this paper, we provide a first systematic study of \covidRelated Android applications. 
We collected from different
channels \covidrelatedApps{} apps that are manually vetted as relevant. 
Then, based on the apps' described goals, our study yields a taxonomy 
of \covidRelated Android apps as a contribution to the literature.   
Our empirical findings reveal that \covidRelated apps have mainly three
purposes: (1) inform users, (2) collect data, and (3) provide tooling
capabilities for users.  
After exploring the inner characteristics (e.g., libraries, permissions, size) and the results 
of security and privacy scanners, we provide first insights into the nature of \covidRelated apps. 
Overall, our empirical study constitutes a first milestone towards understanding 
who, what, and how \covidRelated apps are built. 
We expect future work in the community to go in-depth into each of the dimensions that we have explored.

All artifacts are made available online at:
\begin{center}
    \url{https://github.com/Trustworthy-Software/APKCOVID}
\end{center}

%% file: complexity.tex
\section{Description of complexity metrics}
\label{appendix:complexity}

Below we present five complexity metrics initially described by~\citet{chidamber1994metrics}.

\subsection{Weighted Methods per Class (WMC)}

This metric represents the complexity of individual classes by counting the number of methods per class.
Its purpose is to estimate how challenging the development and maintenance of a class are.
Also, as the value for a given class grows, the impact on a class inheriting it will be greater as they will inherit all the methods.

\subsection{Number Of Children (NOC)}

This metric represents the number of immediate sub-classes of a class in the class hierarchy.
It measures how many sub-classes will inherit the features (fields, methods, etc.) of the parent class.
The greater the number of children, the greater the reuse of properties, meaning that it may require more testing.

\subsection{Lack of Cohesion in Methods (LCOM)}

This metric represents, for each data field in a class, the average percentage of the methods using that field.
It measures the degree of cohesion between methods of a class and data fields of the given class.
The lower the cohesion, the greater the complexity as it increases the probability of errors during the development process.

\subsection{Coupling Between Object classes (CBO)}

This metric represents the dependency that a class has on other classes.
It is computed by counting the number of classes used in a given class.
The greater the value, the greater the complexity since it implies high dependency, hence more testing, and less reusability,

\subsection{Response For a Class (RFC)}

This metric represents the number of unique method calls within a given class.
It is computed by counting all the method invocations in a class only once.
The greater this metric, the greater the complexity since it implies a greater level of understanding for the tester for debugging and/or testing purposes.